\documentclass[fleqn,usenatbib]{mnras}

\usepackage{newtxtext,newtxmath}

\usepackage[T1]{fontenc}
\usepackage{longtable}
\usepackage[justification=centering]{caption}
\DeclareRobustCommand{\VAN}[3]{#2}
\let\VANthebibliography\thebibliography
\def\thebibliography{\DeclareRobustCommand{\VAN}[3]{##3}\VANthebibliography}

\usepackage{graphicx}	
\usepackage{amsmath}	

\title[Chemical models of aminoacetonitrile]{Chemical models of interstellar glycine and adenine precursor aminoacetonitrile (NH$_2$CH$_2$CN)}

\author[X. Zhang et al.]{Xia Zhang,$^{1,2}$
Donghui Quan,$^{3,1}$\thanks{E-mail:donghui.quan@zhejianglab.com}
Xiaohu Li,$^{1,2,4}$\thanks{E-mail:xiaohu.li@xao.ac.cn}
Jarken Esimbek,$^{1,2,4}$
Fangfang Li,$^{1}$
Yan Zhou$^{5}$
\newauthor
and Dalei Li$^{1,2,4}$
\\
$^{1}$ Xinjiang Astronomical Observatory, Chinese Academy of Sciences, 150 Science 1-Street, Urumqi, Xinjiang 830011, China\\
$^{2}$Xinjiang Key Laboratory of Radio Astrophysics, 150 Science1-Street, Urumqi 830011, China\\
$^{3}$ Research Center for Astronomical Computing, Zhejiang Laboratory, Hangzhou 311100, China\\
$^{4}$Key Laboratory of Radio Astronomy, Chinese Academy of Sciences, Nanjing 210008, China\\
$^{5}$BinZhou University, Huanghe Road, Binzhou City, Shandong, 256600, China}

\date{Accepted XXX. Received YYY; in original form ZZZ}

\pubyear{2022}

\begin{document}
\label{firstpage}
\pagerange{\pageref{firstpage}--\pageref{lastpage}}
\maketitle

\begin{abstract}
Aminoacetonitrile (AAN), also known as glycinenitrile, has been suggested as a possible precursor of glycine and adenine in the interstellar medium. Here we present the chemical modeling of AAN and its isomers in hot cores using the three-phase chemical model NAUTILUS with the addition of over 300 chemical reactions of the three AAN isomers and related species. Our models predicted a peak gas phase abundance of AAN reaching the order of 10$^{-8}$, which is consistent with observation towards Sgr B2(N). Regarding the reaction pathways of AAN and its isomers, we found that AAN is primarily formed via free radical reactions on grain surfaces during the early evolutionary stages. Subsequently, it is thermally desorbed into the gas phase as the temperature rises and is then destroyed by positive ions and radicals in gas phase. The isomers of AAN are formed through the hydrogenation reaction of CH$_3$NCN on the grain surface and via electron recombination reactions of ion C$_2$H$_5$N$_2^+$ in gas phase. We speculate that there is a possibility for NCCN and AAN to react with each other, eventually leading to the formation of adenine in hot cores. However, further investigation is required to understand the efficiency of grain surfaces in adenine formation, through theoretical calculations or laboratory experiments in future research.
\end{abstract}




\section{Introduction} \label{sec:intro}
Amino acids, which are the building blocks of proteins and therefore essential for understanding the origin of life, have been discovered in meteorites and comets \citep {glavin2010, elsila2009}. Therefore, it is intriguing to search for amino acids in the interstellar medium (ISM). However, despite a 40-year search, even the simplest amino acid, glycine, has not yet been detected. Aminoacetonitrile (AAN), also known as glycinenitrile, has been proposed as a potential precursor of glycine in the ISM. AAN can form glycine by Strecker synthesis, in which water reacts with a nitrile to form the corresponding acid \citep{peltzer1984}. \citet{choe2023} found AAN can form glycine through a barrierless pathway from the reaction of AAN + OH + H$_2$O with H$_2$O as a catalyst by theoretical calculation and they suggest the reaction can occur on icy grain surfaces in the ISM. \citet{vasconcelos2020} proposed a plausible mechanism for the formation of adenine (C$_5$H$_5$N$_5$), one of the purine of DNA and RNA nucleobases. According to their proposal, AAN may react with cyanogen (NCCN), forming diaminomaleonitrile (C$_4$H$_4$N$_4$), and then C$_4$H$_4$N$_4$ reacts with HCN to produce adenine. 

\begin{figure*}
\centering
\includegraphics[scale=0.4]{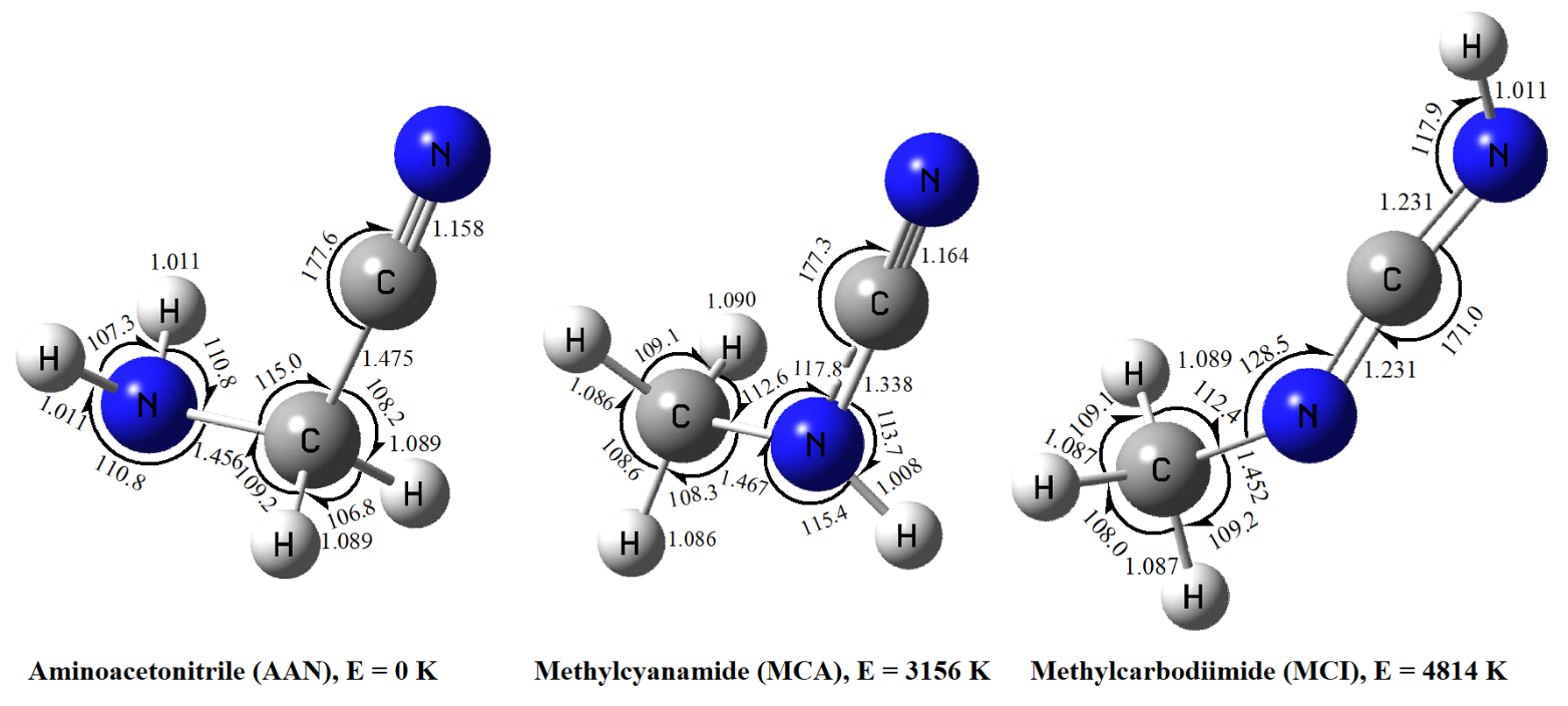}
\caption{Optimized structures of the three isomers. Energies have been calculated at the CCSD(T)/aug-cc-pVTZ//B2PLYPD3/may-cc-pVTZ level. Blue, gray, white spheres correspond to nitrogen, carbon, and hydrogen atoms, respectively.}
\label{fig:molecular_structure}
\end{figure*}

\citet{wirstrom2007} firstly attempted to detect AAN towards four northern hot core sources Orion KL, W51 e1/e2, W3(OH) and S140 using the Onsala 20 m telescope and suggested upper limits to the column density of AAN (1 - 4 $\times$ 10$^{13}$cm$^{-2}$). Later, \citet{belloche2008} successfully detected AAN towards Sgr B2(N) regions using the IRAM 30 m telescope. They speculated that the column density of AAN was 2.8 $\times$ 10$^{16}$cm$^{-2}$ at the temperature of 100 K. In Sgr B2(M) region, they derived an upper limit of 6 $\times$ 10$^{15}$cm$^{-2}$. \citet{richard2018} detected AAN towards Sgr B2(N2) hot core in the EMoCA (Exploring Molecular Complexity with ALMA) survey. They derived a column density of 9.7 $\times$ 10$^{16}$cm$^{-2}$ at a rotational temperature of 150 K for AAN. Additionally, \citet{richard2018} also tentatively detected aminopropionitrile (APN), the family of AAN, which is the next stage in complexity with one additional CH$_2$ group comparing to AAN. The molecule has two structural isomers, 3-aminopropionitrile (NH$_2$CH$_2$CH$_2$CN) and 2-aminopropionitrile (CH$_3$CH(NH$_2$)CN). However, neither 3-aminopropionitrile nor 2-aminopropionitrile was detected. Recently, \citet{melosso2020} used the imaging spectral line survey ReMoCA (Re-exploring Molecular Complexity with ALMA) to search AAN towards Sgr B2(N1) and derived a column density of 1.1 $\times$ 10$^{17}$cm$^{-2}$ at a rotational temperature of 200 K. Besides, AAN was also tentatively detected in the hot molecular core G10.47+0.03 using ALMA data \citep{manna2022, mondal2023}.

AAN is one of nitriles that have been found to exhibit greater stability than glycine acids when exposed to UV photolysis in the solid state at 15 K, according to experiments conducted by \citet{bernstein2004}. These findings suggest that nitriles may be more abundant than acids in the ISM, in part due to their slower  destruction through photolysis. \citet{pelc2016} concluded that AAN was less susceptible to electron capture through low anion yields using experiments and quantum chemical calculations. Therefore, AAN can be considered to exhibit a relatively long lifetime under typical conditions in ISM.

According to \citet{lattelais2010}, C$_2$H$_4$N$_2$ has seventeen structures. AAN is the most stable structure, followed by methylcyanamide (CH$_3$NHCN, MCA) and methylcarbodiimide (CH$_3$NCNH, MCI). The structural images of three isomers are shown in Fig.\ref{fig:molecular_structure}. Among them, MCA and MCI, have not been studied extensively nor been discovered in the ISM \citep{bak1980, sleiman2018, wang2019}. It is worth noting that both MCA and MCI are chiral molecules, meaning that they both have mirror image molecules. Many life-related molecules are chiral, such as amino acids and sugar, where amino acids are always left-handed, and sugars are always right-handed. The first interstellar chiral molecule, propylene oxide (CH$_3$CHOCH$_2$), was discovered in the star formation region Sgr B2 \citep{mcguire2016}. This finding suggests that the chiral molecules essential for life may have originated from space, particularly from star-forming regions. Furthermore, MCA may undergo hydrolysis, similar to AAN, resulting in the formation of methylcarbamic acid (CH$_3$NHCOOH) acid. This acid is an isomer of glycine and is considered the most stable structure \citep{lattelais2011}. 

There are multiple pathways to form AAN. The reaction between methanimine (CH$_2$NH) and HCN (or HNC) was proposed both in gas phase and on surface by \citet{xu2007} and \citet{koch2008} based on theoretical calculation. They discovered that this pathway in the gas phase exhibits significant barriers, whereas they appear to be feasible on the grain surface.

\citet{belloche2009} proposed two grain-phase reactions related to AAN. These reactions involve the interaction between CN and CH$_2$NH, as well as reaction between NH and cyanomethyl radical (H$_2$CCN). Both of them firstly produce NHCH$_2$CN, which subsequently reacts with atomic H to produce AAN, as follows:
\protect\\

\noindent CN + CH$_2$NH $\rightarrow$ NHCH$_2$CN,  \hfill(1) \protect\\

\noindent NH + H$_2$CCN $\rightarrow$ NHCH$_2$CN,  \hfill(2)\protect\\

\noindent NHCH$_2$CN + H $\rightarrow$ NH$_2$CH$_2$CN. \hfill(3)\protect\\

\noindent \citet{belloche2009} also proposed two direct routes to form AAN on the grain surface, they are as follows:
\protect\\

\noindent CN + CH$_2$NH$_2$ $\rightarrow$ NH$_2$CH$_2$CN,  \hfill(4) \protect\\

\noindent NH$_2$ + H$_2$CCN $\rightarrow$ NH$_2$CH$_2$CN. \hfill(5)\protect\\

\noindent Later, \citet{danger2011} demonstrated that AAN can be formed at 20 K from the Vacuum Ultraviolet (VUV) irradiation of acetonitrile (CH$_3$CN) with the presence of ammonia using experimental investigation, which can produce the radicals H$_2$CCN and NH$_2$. Then Reaction (5) subsequently occurs to produce AAN. The reaction between CN and methylamine (CH$_3$NH$_2$) can form AAN, but it has a relatively high barrier \citep{sleiman2018, puzzarini2020}.

For isomeric MCA, \citet{sleiman2018} proposed that the reaction between CN + CH$_3$NH$_2$ can form MCA using experimental and theoretical simulations  at temperatures ranging from 23 to 297 K, which represents the second most significant channel among three. However, \citet{puzzarini2020} found there is high energy barrier by quantum-chemical study. 

So far, three research groups have reported modeling of interstellar AAN \citep{belloche2009, garrod2013, mondal2023}. \citet{belloche2009} modeled the chemistry of Sgr B2(N) using gas-grain-mantel three phase code with addition of reactions of ethyl format, ethyl, n-propyl cyanide and AAN. Later, \citet{garrod2013} used a three-phase model of hot cores to model the formation of glycine, which included the chemistry of AAN from \citet{belloche2009}. \citet{garrod2013} found that the major formation of AAN is the reaction between NH$_2$ and H$_2$CCN (Reaction (5)) and also derived the formation of AAN associated with the thermal desorption of CH$_3$CN, which can produce the radical H$_2$CCN through the gas phase destruction routes. The H$_2$CCN could re-accrete onto the grains and react with NH$_2$ to form AAN. The two studies both produced fractional abundances of AAN on the order of 10$^{-8}$. \citet{mondal2023} also used a three-phase version of NAUTILUS gas-grain code to simulate some nitrogen-bearing complex organic molecules in the hot molecular core, G10.47+0.03. Their chemical network of AAN is based on \citet{belloche2009}. However, these studies did not focus on AAN. So the reactions presented may be incomplete. Additionally, the isomers of AAN were not considered. Therefore, more comprehensive studies are needed to better understand AAN as well as its isomers.

In this study, we consider reactions that have been experimentally approved or theoretically suggested for AAN, MCA, and MCI. Additionally, we also propose new reactive routes which are calculated by quantum chemistry calculations, and present them in Section \ref{sect:chem}. In Section \ref{sect:model}, we show chemical kinetics models that are developed to simulate the formation and destruction mechanisms of the three prebiotic molecules under the interstellar physical conditions. We also explore the possible interstellar environment for further formation of these molecules. Moreover, we conducted an investigation and analysis of the modeling results for cyanamide (NH$_2$CN), methylenimine (CH$_2$NH), methylamine (CH$_3$NH$_2$), cyanomethyl radical (H$_2$CCN), acetonitrile (CH$_3$CN) and cyanogen (NCCN) and discuss the results in Section \ref{sect:result}. Finally, we draw conclusions and summarize our findings in Section \ref{sect:conclu}.

\section{Chemistry of AAN, MCA and MCI }
\label{sect:chem}
\subsection{Formation and destruction of AAN, MCA and MCI}
\begin{figure*}
\centering
\includegraphics[scale=0.68]{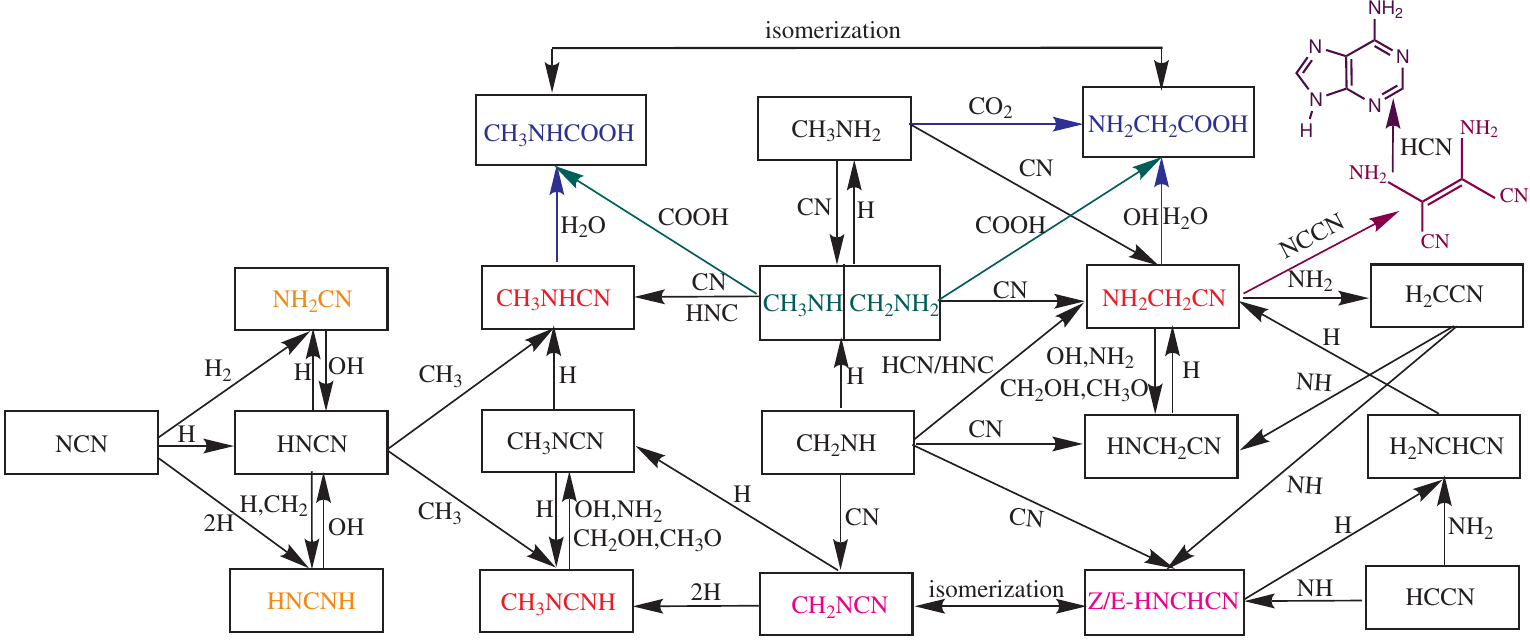}
\caption{The neutral-neutral reactions linking the AAN and its isomers, with species of the same color indicating isomers. Blue arrows are for hydrolysis reactions. Green arrows denote the reactions for the formation of glycine and its isomer on the grain surfaces. Purple and dark purple arrows and corresponding species represent these reactions need to be clarified by theoretical and experimental studies. All these reactions are currently not included in our network.}
\label{fig:AAN}
\end{figure*}

In addition to the previously mentioned chemical reactions of AAN, we expanded our investigation to include other types of reactions involving AAN, as well as reactions concerning MCA and MCI. The major types of reactions considered in this study include cosmic-ray induced photodissociation, UV photodissociation, dissociative recombination, ion-neutral and neutral-neutral reactions. For the first three types of reactions, rate parameters are estimated based on similar reactions already exiting in the chemical network. For ion-molecule reactions, since the branching ratios to these different channels are frequently unmeasured, we assume that reactions proceed equally through different channels, i.e., the branching ratios adopt the same value. And the corresponding rate parameters are calculated using the method developed by \citet{woon2009}. For neutral-neutral reactions, quantum chemical calculations were performed to select only barrierless or lower-barrier exothermic reactions as the possible efficient formation reactions, whose rate coefficients were calculated by the modified Arrhenius equation. The activation energies of the destruction reactions of AAN, by OH, NH$_2$, CH$_2$OH, and CH$_3$O, were estimated based on analogous reactions outlined in \citet{garrod2013}. We then performed quantum chemical calculations to obtain accurate values. Additionally, we also added the destruction of MCA and MCI by the four radicals listed above. Tables \ref{tab:gas-phase} and \ref{tab:activation-energy} present the rate coefficients for gas-phase reactions of AAN, MCA, MCI, and related species, as well as the activation energy barriers for surface/mantle reactions. Moreover, Table \ref{tab:binding-energy} provides the binding energies of selected species utilized in the models. Fig. \ref{fig:AAN} presents the neutral-neutral reactions linking AAN and its isomers, using the of color code indicating the isomers. In addition, blue arrows represent hydrolysis reactions or the reaction among AAN, OH and H$_2$O. Green arrows denote the reactions for the formation of glycine and its isomer on the grain surface, purple and dark purple arrows and corresponding species represent reactions that need to be clarified by theoretical and experimental studies. The reaction pathways of glycine and its isomer, C$_4$H$_4$N$_4$ and adenine are currently not included in this work.

\subsection{Quantum chemical calculations}
All quantum chemical calculations were performed using the GAUSSIAN 16 program \citep{Gaussian16}. The structures of the reactants, transition states, intermediates, and products were optimized, and zero-point energy corrections were determined using the double-hybrid B2PLYP functions \citep{grimme2006} with the may-cc-pVTZ basis set. Semi-empirical dispersion contributions were also included by using the D3BJ model \citep{goerigk2011, grimme2011}. Single-point energies were then calculated using the highly cost-effective method of CCSD(T) \citep{purvis1982, scuseria1988, scuseria1989}, together with the basis set of aug-cc-pVTZ. Saddle points were assigned to reaction paths by using the intrinsic reaction coordinate (IRC) \citep{fukui1981} calculations at the same B2PLYP level as the identification of reactants and products. 
\clearpage
\onecolumn
\begin{longtable}
{p{8cm}p{1.4cm}p{1.4cm}p{1.4cm}p{1.2cm}}
\caption{Summary of the rate coefficients of gas-phase reactions involving AAN, MCA, MCI and related species.}
 \label{tab:gas-phase}\\
  \hline
  \endfirsthead
 \multicolumn{5}{c}{Continuation of Table \ref{tab:gas-phase}}\\
 \hline
 Reaction    & $\alpha$     & $\beta$        & $\gamma$    & Ref.\\
 \hline
 \endhead
 \hline
 \endfoot
 \multicolumn{5}{l}{$^a$ Estimation according to analogous reaction rate coefficients in the OSU and KIDA network (\citet{hasegawa1992};}\\
 \multicolumn{5}{l}{\citet{garrod2007}; \citet{hassel2008}; \citet{quan2010}; \citet{wakelam2015}).}\\
 \multicolumn{5}{l}{$^b$ Rate coefficients for unmeasured reactions between ions and neutral species with a dipole moment are computed using}\\
 \multicolumn{5}{l}{the Su-Chesnavich capture approach. This approach is discussed in \citet{su1982}; \citet{woon2009}.}\\
\multicolumn{5}{l}{$^c$ \citet{puzzarini2020}.}\\
\multicolumn{5}{l}{$^d$ This work, based on potential and barrier calculations.}
 \endlastfoot
  Reaction                                            & $\alpha$      & $\beta$   & $\gamma$ & Ref.\\
  \hline
  Cosmic Ray Induced Photodissociation                &(-)            &(-)        &(-)       & \\
  \hline
  NH$_2$CH$_2$CN $\rightarrow$ NH$_2$ + H$_2$CCN      & 1.50E+03	    & 0	        & 0	       & a\\
  NH$_2$CH$_2$CN $\rightarrow$ CH$_2$NH$_2$ + CN      & 1.50E+03	    & 0	        & 0	       & a\\
  NH$_2$CH$_2$CN $\rightarrow$ NH$_2$CHCN + H         & 1.50E+03	    & 0	        & 0	       & a\\
  NH$_2$CH$_2$CN $\rightarrow$ NHCH$_2$CN + H         & 1.50E+03	    & 0	        & 0	       & a\\
  NHCH$_2$CN $\rightarrow$ H$_2$CCN + NH              & 1.50E+03	    & 0	        & 0	       & a\\
  NHCH$_2$CN $\rightarrow$ CH$_2$NH + CN              & 1.50E+03	    & 0	        & 0	       & a\\
  NH$_2$CHCN $\rightarrow$ HCCN + NH$_2$              & 1.50E+03	    & 0	        & 0	       & a\\
  NH$_2$CHCN $\rightarrow$ CH$_2$NH + CN              & 1.50E+03	    & 0	        & 0	       & a\\
  CH$_3$NHCN $\rightarrow$ CH$_3$NH + CN              & 1.50E+03	    & 0	        & 0	       & a\\
  CH$_3$NHCN $\rightarrow$ CH$_3$ + HNCN              & 1.50E+03	    & 0	        & 0	       & a\\
  CH$_3$NHCN $\rightarrow$ CH$_3$NCN + H              & 1.50E+03	    & 0	        & 0	       & a\\
  CH$_3$NHCN $\rightarrow$ CH$_2$NHCN + H             & 1.50E+03	    & 0	        & 0	       & a\\
  CH$_3$NCNH $\rightarrow$ CH$_3$ + HNCN              & 1.50E+03	    & 0	        & 0	       & a\\
  CH$_3$NCNH $\rightarrow$ CH$_3$NCN + H              & 1.50E+03	    & 0	        & 0	       & a\\
  CH$_3$NCNH $\rightarrow$ CH$_2$NCNH + H             & 1.50E+03	    & 0	        & 0	       & a\\
  CH$_3$NCN $\rightarrow$ CH$_3$ + NCN                & 1.50E+03	    & 0	        & 0	       & a\\
  CH$_2$NCNH $\rightarrow$ CH$_2$ + HNCN              & 1.50E+03	    & 0	        & 0	       & a\\
  CH$_2$NHCN $\rightarrow$ CH$_2$ + HNCN              & 1.50E+03	    & 0	        & 0	       & a\\
  CH$_2$NHCN $\rightarrow$ CH$_2$NH + CN              & 1.50E+03	    & 0	        & 0	       & a\\
  CH$_3$NH $\rightarrow$ CH$_2$NH + H                 & 1.50E+03	    & 0	        & 0	       & a\\
  CH$_3$NH $\rightarrow$ CH$_3$ + NH                  & 1.50E+03	    & 0	        & 0	       & a\\
  \hline
  Photodissociation                                   & $(s^{-1})$    &(-)        &(-)       &\\
  \hline
  NH$_2$CH$_2$CN $\rightarrow$ NH$_2$ + H$_2$CCN      & 1.00E-09	    & 0	        & 1.9	   & a\\
  NH$_2$CH$_2$CN $\rightarrow$ CH$_2$NH$_2$ + CN      & 1.00E-09	    & 0	        & 1.9      & a\\
  NH$_2$CH$_2$CN $\rightarrow$ NH$_2$CHCN + H         & 1.00E-09	    & 0	        & 1.9	   & a\\
  NH$_2$CH$_2$CN $\rightarrow$ NHCH$_2$CN + H         & 1.00E-09	    & 0	        & 1.9	   & a\\
  NHCH$_2$CN $\rightarrow$ H$_2$CCN + NH              & 1.00E-09	    & 0	        & 1.9      & a\\
  NHCH$_2$CN $\rightarrow$ CH$_2$NH + CN              & 1.00E-09	    & 0	        & 1.9	   & a\\
  NH$_2$CHCN $\rightarrow$ HCCN + NH$_2$              & 1.00E-09	    & 0	        & 1.9	   & a\\
  NH$_2$CHCN $\rightarrow$ CH$_2$NH + CN              & 1.00E-09	    & 0	        & 1.9	   & a\\
  CH$_3$NHCN $\rightarrow$ CH$_3$NH + CN              & 1.00E-09	    & 0	        & 1.9	   & a\\
  CH$_3$NHCN $\rightarrow$ CH$_3$ + HNCN              & 1.00E-09	    & 0	        & 1.9	   & a\\
  CH$_3$NHCN $\rightarrow$ CH$_3$NCN + H              & 1.00E-09	    & 0	        & 1.9	   & a\\
  CH$_3$NHCN $\rightarrow$ CH$_2$NHCN + H             & 1.00E-09	    & 0	        & 1.9	   & a\\
  CH$_3$NCNH $\rightarrow$ CH$_3$ + HNCN              & 1.00E-09	    & 0	        & 1.9	   & a\\
  CH$_3$NCNH $\rightarrow$ CH$_3$NCN + H              & 1.00E-09	    & 0	        & 1.9	   & a\\
  CH$_3$NCNH $\rightarrow$ CH$_2$NCNH + H             & 1.00E-09	    & 0	        & 1.9	   & a\\
  CH$_3$NCN $\rightarrow$ CH$_3$ + NCN                & 1.00E-09	    & 0	        & 1.9	   & a\\
  CH$_2$NCNH $\rightarrow$ CH$_2$ + HNCN              & 1.00E-09	    & 0	        & 1.9	   & a\\
  CH$_2$NHCN $\rightarrow$ CH$_2$ + HNCN              & 1.00E-09	    & 0	        & 1.9	   & a\\
  CH$_2$NHCN $\rightarrow$ CH$_2$NH + CN              & 1.00E-09	    & 0	        & 1.9	   & a\\
  CH$_3$NH $\rightarrow$ CH$_2$NH + H                 & 1.00E-09	    & 0	        & 1.9	   & a\\
  CH$_3$NH $\rightarrow$ CH$_3$ + NH                  & 1.00E-09	    & 0	        & 1.9	   & a\\
  \hline
  Ion-neutral                                                    &(-)      &$(cm^3 s^{-1})$ &(-)    &\\
  \hline
  C$^+$ + NH$_2$CH$_2$CN $\rightarrow$ C$_2$H$_4$N$_2^+$ + C     & 0.25    &1.75E-09        & 4.09  & b\\
  C$^+$ + NH$_2$CH$_2$CN $\rightarrow$ C$_2$N$^+$ + CH$_2$NH$_2$ & 0.25    &1.75E-09        & 4.09  & b\\
  C$^+$ + NH$_2$CH$_2$CN $\rightarrow$ H$_2$NC$^+$ + H$_2$CCN    & 0.25    &1.75E-09        & 4.09  & b\\
  C$^+$ + NH$_2$CH$_2$CN $\rightarrow$ HC$_3$NH$^+$ + NH$_2$     & 0.25    &1.75E-09        & 4.09  & b\\
  H$^+$ + NH$_2$CH$_2$CN $\rightarrow$ C$_2$H$_4$N$_2^+$ + H     & 0.20    &5.54E-09        & 4.09  & b\\
  H$^+$ + NH$_2$CH$_2$CN $\rightarrow$ HCN$^+$ + CH$_2$NH$_2$    & 0.20    &5.54E-09        & 4.09  & b\\
  H$^+$ + NH$_2$CH$_2$CN $\rightarrow$ CH$_3$NH$_2^+$ + CN       & 0.20    &5.54E-09        & 4.09  & b\\
  H$^+$ + NH$_2$CH$_2$CN $\rightarrow$ NH$_3^+$ + H$_2$CCN       & 0.20    &5.54E-09        & 4.09  & b\\
  H$^+$ + NH$_2$CH$_2$CN $\rightarrow$ CH$_3$CN$^+$ + NH$_2$     & 0.20    &5.54E-09        & 4.09  & b\\
  He$^+$ + NH$_2$CH$_2$CN $\rightarrow$ CH$_2$NH$_2^+$ + CN + He       & 0.25    &2.84E-09        & 4.09  & b\\
  He$^+$ + NH$_2$CH$_2$CN $\rightarrow$ CH$_2$NH$_2$ + CN$^+$ + He     & 0.25    &2.84E-09        & 4.09  & b\\
  He$^+$ + NH$_2$CH$_2$CN $\rightarrow$ CH$_2$CN$^+$ + NH$_2$ + He     & 0.25    &2.84E-09        & 4.09  & b\\
  He$^+$ + NH$_2$CH$_2$CN $\rightarrow$ H$_2$CCN + NH$_2^+$ + He       & 0.25    &2.84E-09        & 4.09  & b\\
  H$_3^+$ + NH$_2$CH$_2$CN $\rightarrow$ C$_2$H$_5$N$_2^+$ + H$_2$     & 0.20    &3.25E-09        & 4.09  & b\\       
  H$_3^+$ + NH$_2$CH$_2$CN $\rightarrow$ NH$_3^+$ + H$_2$CCN + H$_2$   & 0.20    &3.25E-09        & 4.09  & b\\
  H$_3^+$ + NH$_2$CH$_2$CN $\rightarrow$ NH$_2$ + CH$_3$CN$^+$ + H$_2$ & 0.20    &3.25E-09        & 4.09  & b\\
  H$_3^+$ + NH$_2$CH$_2$CN $\rightarrow$ CH$_2$NH$_2$ + HCN$^+$ + H$_2$ & 0.20    &3.25E-09        & 4.09  & b\\
  H$_3^+$ + NH$_2$CH$_2$CN $\rightarrow$ CH$_3$NH$_2^+$ + CN + H$_2$    & 0.20    &3.25E-09        & 4.09  & b\\
  H$_3$O$^+$ + NH$_2$CH$_2$CN $\rightarrow$ C$_2$H$_5$N$_2^+$ + H$_2$O  & 0.20    &1.46E-09        & 4.09  & b\\
  H$_3$O$^+$ + NH$_2$CH$_2$CN $\rightarrow$ NH$_3^+$ + H$_2$CCN + H$_2$O    & 0.20    &1.46E-09        & 4.09  & b\\
  H$_3$O$^+$ + NH$_2$CH$_2$CN $\rightarrow$ NH$_2$ + CH$_3$CN$^+$ + H$_2$O  & 0.20    &1.46E-09        & 4.09  & b\\
  H$_3$O$^+$ + NH$_2$CH$_2$CN $\rightarrow$ CH$_2$NH$_2$ + HCN$^+$ + H$_2$O & 0.20    &1.46E-09        & 4.09  & b\\
  H$_3$O$^+$ + NH$_2$CH$_2$CN $\rightarrow$ CH$_3$NH$_2$ + CN + H$_2$O      & 0.20    &1.46E-09        & 4.09  & b\\
  HCO$^+$ + NH$_2$CH$_2$CN $\rightarrow$ C$_2$H$_5$N$_2^+$ + CO      & 0.20    &1.26E-09        & 4.09  & b\\
  HCO$^+$ + NH$_2$CH$_2$CN $\rightarrow$ NH$_3^+$ + H$_2$CCN + CO    & 0.20    &1.26E-09        & 4.09  & b\\
  HCO$^+$ + NH$_2$CH$_2$CN $\rightarrow$ NH$_2$ + CH$_3$CN$^+$ + CO  & 0.20    &1.26E-09        & 4.09  & b\\
  HCO$^+$ + NH$_2$CH$_2$CN $\rightarrow$ CH$_2$NH$_2$ + HCN$^+$ + CO & 0.20    &1.26E-09        & 4.09  & b\\
  HCO$^+$ + NH$_2$CH$_2$CN $\rightarrow$ CH$_3$NH$_2^+$ + CN + CO    & 0.20    &1.26E-09        & 4.09  & b\\
  C$^+$ + CH$_3$NHCN $\rightarrow$ C$_2$H$_4$N$_2^+$ + C     & 0.25    &1.81E-09        & 7.05  & b\\
  C$^+$ + CH$_3$NHCN $\rightarrow$ C$_2$N$^+$ + CH$_3$NH     & 0.25    &1.81E-09        & 7.05  & b\\
  C$^+$ + CH$_3$NHCN $\rightarrow$ CHNCN$^+$ + CH$_3$        & 0.25    &1.81E-09        & 7.05  & b\\
  C$^+$ + CH$_3$NHCN $\rightarrow$ HNCN + C$_2$H$_3^+$       & 0.25    &1.81E-09        & 7.05  & b\\
  H$^+$ + CH$_3$NHCN $\rightarrow$ C$_2$H$_4$N$_2^+$ + H     & 0.20    &5.73E-09        & 7.05  & b\\
  H$^+$ + CH$_3$NHCN $\rightarrow$ HCN$^+$ + CH$_3$NH        & 0.20    &5.73E-09        & 7.05  & b\\
  H$^+$ + CH$_3$NHCN $\rightarrow$ CH$_3$NH$_2^+$ + CN       & 0.20    &5.73E-09        & 7.05  & b\\
  H$^+$ + CH$_3$NHCN $\rightarrow$ CH$_4^+$ + HNCN           & 0.20    &5.73E-09        & 7.05  & b\\
  H$^+$ + CH$_3$NHCN $\rightarrow$ CH$_3$ + NH$_2$CN$^+$     & 0.20    &5.73E-09        & 7.05  & b\\
  He$^+$ + CH$_3$NHCN $\rightarrow$ CH$_3$NH$^+$ + CN + He   & 0.25    &2.94E-09        & 7.05  & b\\
  He$^+$ + CH$_3$NHCN $\rightarrow$ CH$_3$NH + CN$^+$ + He   & 0.25    &2.94E-09        & 7.05  & b\\
  He$^+$ + CH$_3$NHCN $\rightarrow$ HNCN$^+$ + CH$_3$ + He   & 0.25    &2.94E-09        & 7.05  & b\\
  He$^+$ + CH$_3$NHCN $\rightarrow$ HNCN + CH$_3^+$ + He     & 0.25    &2.94E-09        & 7.05  & b\\
  H$_3^+$ + CH$_3$NHCN $\rightarrow$ C$_2$H$_5$N$_2^+$ + H$_2$   & 0.20    &3.37E-09        & 7.05  & b\\
  H$_3^+$ + CH$_3$NHCN $\rightarrow$ HCN$^+$ + CH$_3$NH + H$_2$  & 0.20    &3.37E-09        & 7.05  & b\\
  H$_3^+$ + CH$_3$NHCN $\rightarrow$ CN + CH$_3$NH$_2^+$ + H$_2$ & 0.20    &3.37E-09        & 7.05  & b\\
  H$_3^+$ + CH$_3$NHCN $\rightarrow$ CH$_4^+$ + HNCN + H$_2$     & 0.20    &3.37E-09        & 7.05  & b\\
  H$_3^+$ + CH$_3$NHCN $\rightarrow$ CH$_3$ + NH$_2$CN$^+$ + H$_2$  & 0.20    &3.37E-09        & 7.05  & b\\
  H$_3$O$^+$ + CH$_3$NHCN $\rightarrow$ C$_2$H$_5$N$_2^+$ + H$_2$O  & 0.20    &1.51E-09        & 7.05  & b\\
  H$_3$O$^+$ + CH$_3$NHCN $\rightarrow$ HCN$^+$ + CH$_3$NH + H$_2$O & 0.20    &1.51E-09        & 7.05  & b\\
  H$_3$O$^+$ + CH$_3$NHCN $\rightarrow$ CN + CH$_3$NH$_2^+$ + H$_2$O & 0.20    &1.51E-09        & 7.05  & b\\
  H$_3$O$^+$ + CH$_3$NHCN $\rightarrow$ CH$_4^+$ + HNCN + H$_2$O     & 0.20    &1.51E-09        & 7.05  & b\\
  H$_3$O$^+$ + CH$_3$NHCN $\rightarrow$ CH$_3$ + NH$_2$CN$^+$ + H$_2$O & 0.20    &1.51E-09        & 7.05  & b\\
  HCO$^+$ + CH$_3$NHCN $\rightarrow$ C$_2$H$_5$N$_2^+$ + CO      & 0.20    &1.30E-09        & 7.05  & b\\
  HCO$^+$ + CH$_3$NHCN $\rightarrow$ HCN$^+$ + CH$_3$NH + CO     & 0.20    &1.30E-09        & 7.05  & b\\
  HCO$^+$ + CH$_3$NHCN $\rightarrow$ CN + CH$_3$NH$_2^+$ + CO    & 0.20    &1.30E-09        & 7.05  & b\\
  HCO$^+$ + CH$_3$NHCN $\rightarrow$ CH$_4^+$ + HNCN + CO        & 0.20    &1.30E-09        & 7.05  & b\\
  HCO$^+$ + CH$_3$NHCN $\rightarrow$ CH$_3$ + NH$_2$CN$^+$ + CO  & 0.20    &1.30E-09        & 7.05  & b\\
  C$^+$ + CH$_3$NCNH $\rightarrow$ C$_2$H$_4$N$_2^+$ + C     & 0.30    &1.87E-09        & 3.05  & b\\
  C$^+$ + CH$_3$NCNH $\rightarrow$ HNCN + C$_2$H$_3^+$       & 0.35    &1.87E-09        & 3.05  & b\\
  C$^+$ + CH$_3$NCNH $\rightarrow$ CHNCN$^+$ + CH$_3$        & 0.35    &1.87E-09        & 3.05  & b\\
  H$^+$ + CH$_3$NCNH $\rightarrow$ C$_2$H$_4$N$_2^+$ + H     & 0.30    &5.92E-09        & 3.05  & b\\
  H$^+$ + CH$_3$NCNH $\rightarrow$ CH$_4^+$ + HNCN           & 0.35    &5.92E-09        & 3.05  & b\\
  H$^+$ + CH$_3$NCNH $\rightarrow$ CH$_3$ + NH$_2$CN$^+$     & 0.35    &5.92E-09        & 3.05  & b\\
  He$^+$ + CH$_3$NCNH $\rightarrow$ HNCN$^+$ + CH$_3$ + He   & 0.50    &3.03E-09        & 3.05  & b\\
  He$^+$ + CH$_3$NCNH $\rightarrow$ HNCN + CH$_3^+$ + He     & 0.50    &3.03E-09        & 3.05  & b\\
  H$_3^+$ + CH$_3$NCNH $\rightarrow$ C$_2$H$_5$N$_2^+$ + H$_2$ & 0.50    &3.47E-09        & 3.05  & b\\
  H$_3^+$ + CH$_3$NCNH $\rightarrow$ CH$_4^+$ + HNCN + H$_2$   & 0.50    &3.47E-09        & 3.05  & b\\
  H$_3$O$^+$ + CH$_3$NCNH $\rightarrow$ C$_2$H$_5$N$_2^+$ + H$_2$O  & 0.50    &1.56E-09        & 3.05  & b\\
  H$_3$O$^+$ + CH$_3$NCNH $\rightarrow$ CH$_4^+$ + HNCN + H$_2$O    & 0.50    &1.56E-09        & 3.05  & b\\
  HCO$^+$ + CH$_3$NCNH $\rightarrow$ C$_2$H$_5$N$_2^+$ + CO         & 0.50    &1.34E-09        & 3.05  & b\\
  HCO$^+$ + CH$_3$NCNH $\rightarrow$ CH$_4^+$ + HNCN + CO           & 0.50    &1.34E-09        & 3.05  & b\\
  \hline
  Neutral-neutral                                 & $(cm^3 s^{-1})$  &(-) &(-)       &\\
  \hline
  CN + CH$_3$NH$_2$ $\rightarrow$ CH$_3$NH + HCN      & 3.94E-10	& 0.0	 & 0.00E+00	   & c\\
  CN + CH$_3$NH$_2$ $\rightarrow$ CH$_2$NH$_2$ + HCN  & 8.87E-10	& 0.0	 & 0.00E+00	   & c\\
  CN + CH$_3$NH$_2$ $\rightarrow$ NH$_2$CH$_2$CN + H  & 1.73E-17	& 0.0	 & 0.00E+00	   & c\\
  HNC + CH$_3$NH $\rightarrow$ CH$_3$NHCN + H         & 8.70E-17	& 0.5	 & 3.49E+03	   & d\\
  OH + NH$_2$CH$_2$CN $\rightarrow$ NHCH$_2$CN + H$_2$O          & 4.31E-10	& 0.5	 & 1.11E+03	   & d\\
  NH$_2$ + NH$_2$CH$_2$CN $\rightarrow$ NHCH$_2$CN + NH$_3$      & 4.69E-10	& 0.5	 & 3.77E+03	   & d\\
  CH$_2$OH + NH$_2$CH$_2$CN $\rightarrow$ NHCH$_2$CN + CH$_3$OH  & 4.48E-10	& 0.5	 & 5.10E+03	   & d\\
  CH$_3$O + NH$_2$CH$_2$CN $\rightarrow$ NHCH$_2$CN + CH$_3$OH   & 4.28E-10	& 0.5	 & 3.05E+03	   & d\\
  OH + CH$_3$NHCN $\rightarrow$ CH$_2$NHCN + H$_2$O         & 5.05E-10	& 0.5	 & 4.30E+02	   & d\\
  NH$_2$ + CH$_3$NHCN $\rightarrow$ CH$_2$NHCN + NH$_3$     & 5.48E-10	& 0.5	 & 4.01E+03	   & d\\
  CH$_2$OH + CH$_3$NHCN $\rightarrow$ CH$_2$NHCN + CH$_3$OH & 5.16E-10	& 0.5	 & 4.47E+03	   & d\\
  CH$_3$O + CH$_3$NHCN $\rightarrow$ CH$_2$NHCN + CH$_3$OH  & 4.94E-10	& 0.5	 & 2.78E+03	   & d\\
  OH + CH$_3$NCNH $\rightarrow$ CH$_2$NCNH + H$_2$O         & 5.28E-10	& 0.5	 & 7.07E+02	   & d\\
  NH$_2$ + CH$_3$NCNH $\rightarrow$ CH$_2$NCNH + NH$_3$     & 5.72E-10	& 0.5	 & 4.42E+03	   & d\\
  CH$_2$OH + CH$_3$NCNH $\rightarrow$ CH$_2$NCNH + CH$_3$OH & 5.37E-10	& 0.5	 & 4.47E+03	   & d\\
  CH$_3$O + CH$_3$NCNH $\rightarrow$ CH$_2$NCNH + CH$_3$OH  & 5.15E-10	& 0.5	 & 2.99E+03	   & d\\
  OH + CH$_3$NCNH $\rightarrow$ CH$_3$NCN + H$_2$O          & 5.28E-10	& 0.5	 & 0.00E+00	   & d\\
  NH$_2$ + CH$_3$NCNH $\rightarrow$ CH$_3$NCN + NH$_3$      & 5.72E-10	& 0.5	 & 1.74E+03	   & d\\
  CH$_2$OH + CH$_3$NCNH $\rightarrow$ CH$_3$NCN + CH$_3$OH  & 5.37E-10	& 0.5	 & 2.49E+03	   & d\\
  CH$_3$O + CH$_3$NCNH $\rightarrow$ CH$_3$NCN + CH$_3$OH   & 5.15E-10	& 0.5	 & 9.55E+02	   & d\\
  \hline
  Dissociative Recombination                          & $(cm^3 s^{-1})$  &(-)   &(-) &\\
  \hline
  C$_2$H$_4$N$_2^+$ + e$^-$ $\rightarrow$ CH$_2$NH$_2$ + CN    & 1.50E-07	  & -0.5	& 0	  & a\\
  C$_2$H$_4$N$_2^+$ + e$^-$ $\rightarrow$ H$_2$CCN + NH$_2$    & 1.50E-07	  & -0.5	& 0	  & a\\
  C$_2$H$_4$N$_2^+$ + e$^-$ $\rightarrow$ CH$_3$NH + CN        & 1.50E-07	  & -0.5	& 0	  & a\\
  C$_2$H$_4$N$_2^+$ + e$^-$ $\rightarrow$ CH$_3$ + HNCN        & 1.50E-07	  & -0.5	& 0	  & a\\
  C$_2$H$_5$N$_2^+$ + e$^-$ $\rightarrow$ NH$_2$CH$_2$CN + H   & 1.50E-07	  & -0.5	& 0	  & a\\
  C$_2$H$_5$N$_2^+$ + e$^-$ $\rightarrow$ CH$_2$NH$_2$ + HNC   & 1.50E-07	  & -0.5	& 0	  & a\\
  C$_2$H$_5$N$_2^+$ + e$^-$ $\rightarrow$ CH$_3$CN + NH$_2$    & 1.50E-07	  & -0.5	& 0	  & a\\
  C$_2$H$_5$N$_2^+$ + e$^-$ $\rightarrow$ CH$_3$NHCN + H       & 1.50E-07	  & -0.5	& 0	  & a\\
  C$_2$H$_5$N$_2^+$ + e$^-$ $\rightarrow$ CH$_3$NCNH + H       & 1.50E-07	  & -0.5	& 0	  & a\\
  C$_2$H$_5$N$_2^+$ + e$^-$ $\rightarrow$ CH$_3$NH$_2$ + CN    & 1.50E-07	  & -0.5	& 0	  & a\\
  C$_2$H$_5$N$_2^+$ + e$^-$ $\rightarrow$ CH$_3$ + NH$_2$CN    & 1.50E-07	  & -0.5	& 0	  & a\\
  CH$_3$NH$^+$ + e$^-$ $\rightarrow$ CH$_3$ + NH               & 1.50E-07	  & -0.5	& 0	  & a\\ 
\hline
\end{longtable}
\clearpage
\twocolumn

\begin{table*}
 \caption{Activation energy barrier values of important surface/mantle reactions.}
 \label{tab:activation-energy}
 \begin{tabular}{p{8cm}p{4cm}p{1.2cm}}
 \hline
  Surface reactions  & E$_a$(K)      & Ref.\\
  \hline
  CN + CH$_2$NH$_2$ $\rightarrow$ NH$_2$CH$_2$CN  & 0   & a\\
  NH$_2$ + H$_2$CCN $\rightarrow$ NH$_2$CH$_2$CN  & 0   & a\\
  HCN + CH$_2$NH $\rightarrow$ NH$_2$CH$_2$CN  & 3.17E+03   & b\\ 
  HNC + CH$_2$NH $\rightarrow$ NH$_2$CH$_2$CN  & 3.52E+02   & b\\
  NH +  H$_2$CCN $\rightarrow$ NHCH$_2$CN         & 0   & a\\
  CN +  CH$_2$NH $\rightarrow$ NHCH$_2$CN         & 0   & c\\
  NH$_2$ + HCCN $\rightarrow$ NH$_2$CHCN          & 0   & d\\
  E-HNCHCN + H $\rightarrow$ NH$_2$CHCN        & 1.40E+03   & e\\
  Z-HNCHCN + H $\rightarrow$ NH$_2$CHCN        & 1.30E+03   & e\\
  NHCH$_2$CN + H $\rightarrow$ NH$_2$CH$_2$CN     & 0   & a\\
  NH$_2$CHCN + H $\rightarrow$ NH$_2$CH$_2$CN     & 0   & d\\
  H + CH$_2$NH $\rightarrow$ CH$_3$NH             & 0   & f\\
  NH + CH$_3$ $\rightarrow$ CH$_3$NH              & 0   & d\\
  H + CH$_3$NH $\rightarrow$ CH$_3$NH$_2$         & 0   & f\\
  CH$_3$ + HNCN $\rightarrow$ CH$_3$NHCN          & 0   & d\\
  CH$_3$ + HNCN $\rightarrow$ CH$_3$NCNH          & 0   & d\\
  CN + CH$_3$NH $\rightarrow$ CH$_3$NHCN          & 0   & d\\
  HNC + CH$_3$NH $\rightarrow$ CH$_3$NHCN + H     & 3.49E+03   & d\\
  H + CH$_2$NCN $\rightarrow$ CH$_3$NCN           & 1.19E+03   & d\\
  H + CH$_2$NCN $\rightarrow$ CH$_2$NCNH          & 3.71E+03   & d\\
  H + CH$_3$NCN $\rightarrow$ CH$_3$NHCN          & 0  & d\\
  H + CH$_3$NCN $\rightarrow$ CH$_3$NCNH          & 0  & d\\
  H + CH$_2$NCNH $\rightarrow$ CH$_3$NCNH         & 0  & d\\
  OH + NH$_2$CH$_2$CN $\rightarrow$ H$_2$O + NHCH$_2$CN          & 1.11E+03   & g, d\\
  NH$_2$ + NH$_2$CH$_2$CN $\rightarrow$ NH$_3$ + NHCH$_2$CN      & 3.77E+03   & g, d\\
  CH$_2$OH + NH$_2$CH$_2$CN $\rightarrow$ CH$_3$OH + NHCH$_2$CN  & 5.10E+03   & g, d\\   
  CH$_3$O + NH$_2$CH$_2$CN $\rightarrow$ CH$_3$OH + NHCH$_2$CN   & 3.05E+03   & g, d\\
  OH + CH$_3$NHCN $\rightarrow$ CH$_2$NHCN + H$_2$O         & 4.30E+02	   & d\\
  NH$_2$ + CH$_3$NHCN $\rightarrow$ CH$_2$NHCN + NH$_3$     & 4.01E+03	   & d\\
  CH$_2$OH + CH$_3$NHCN $\rightarrow$ CH$_2$NHCN + CH$_3$OH & 4.47E+03	   & d\\
  CH$_3$O + CH$_3$NHCN $\rightarrow$ CH$_2$NHCN + CH$_3$OH  & 2.78E+03	   & d\\
  OH + CH$_3$NCNH $\rightarrow$ CH$_2$NCNH + H$_2$O         & 7.07E+02	   & d\\
  NH$_2$ + CH$_3$NCNH $\rightarrow$ CH$_2$NCNH + NH$_3$     & 4.42E+03	   & d\\
  CH$_2$OH + CH$_3$NCNH $\rightarrow$ CH$_2$NCNH + CH$_3$OH & 4.47E+03	   & d\\
  CH$_3$O + CH$_3$NCNH $\rightarrow$ CH$_2$NCNH + CH$_3$OH  & 2.99E+03	   & d\\
  OH + CH$_3$NCNH $\rightarrow$ CH$_3$NCN + H$_2$O          & 0.00E+00	   & d\\
  NH$_2$ + CH$_3$NCNH $\rightarrow$ CH$_3$NCN + NH$_3$      & 1.74E+03	   & d\\
  CH$_2$OH + CH$_3$NCNH $\rightarrow$ CH$_3$NCN + CH$_3$OH  & 2.49E+03	   & d\\
  CH$_3$O + CH$_3$NCNH $\rightarrow$ CH$_3$NCN + CH$_3$OH   & 9.55E+02	   & d\\
 \hline
 \multicolumn{3}{l}{The dust surface reactions related to AAN, MCA and MCI are shown. Ea represents the value of the activation barrier.}\\
 \multicolumn{3}{l}{Since radical species are so reactive, radical–radical reactions would have no activation barriers.}\\
 \multicolumn{3}{l}{$^a$ \citet{belloche2009}. $^b$ \citet{xu2007, koch2008}. $^c$ \citet{basiuk2001}.}\\
 \multicolumn{3}{l}{$^d$ This work, based on potential and barrier calculations.}\\
 \multicolumn{3}{l}{$^e$ \citet{shingledecker2020}. $^f$ \citet{woon2002,suzuki2016}. $^g$ \citet{garrod2013}.}\\
 \end{tabular}
\end{table*}

\begin{table}
 \caption{Binding energies of involved species.}
 \label{tab:binding-energy}
 \centering
 \begin{tabular}{p{2.5cm}p{2cm}p{2.5cm}}
  \hline
  Species    & E$\rm{_{des}}$(K)    &Ref.\\
  \hline
   NH$_2$CH$_2$CN & 5480	 & \citet{garrod2013}\\
   NHCH$_2$CN     & 5030	 & NH$_2$CH$_2$CN-H\\
   NH$_2$CHCN     & 5030	 & NH$_2$CH$_2$CN-H\\
   CH$_3$NH       & 3414     & \citet{das2018}\\
   CH$_3$NHCN     & 5480	 & NH$_2$CH$_2$CN\\
   CH$_3$NCNH     & 5480	 & NH$_2$CH$_2$CN\\
   CH$_3$NCN      & 5030    & NHCH$_2$CN\\
   CH$_2$NCNH     & 5030    & NHCH$_2$CN\\
   CH$_2$NHCN     & 5030    & NHCH$_2$CN\\
\hline
 \end{tabular}
\end{table}

\section{ASTROCHEMICAL MODEL}
\label{sect:model}
To investigate the formation and destruction mechanism of AAN and its isomers MCA, MCI, we use the three-phase NAUTILUS chemical code \citep{ruaud2016}, which includes the gas, the dust grain surface and the icy mantle. In this study, our gas phase chemistry is based on the public chemical network kida.uva.2014 \citep{wakelam2015}. The network for surface reactions and gas-grain interactions is based on the one from \citet{garrod2007} with several additional processes from \citep{ruaud2015}. And the chemical network has been updated by incorporating the chemistry of HNCO and its metastable isomers, ethanimine (CH$_3$CHNH), cyanomethanimine (E-, Z-HNCHCN, CH$_2$NCN) and cyanamide (NH$_2$CN) \citep{quan2010, quan2016, zhang2020, zhang2023}. Additionally, reactions related to propylene oxide have been integrated based on the work of  \citet{das2019}. Given the significance of radical-radical chemical reactions in the formation of Complex Organic Molecules (COMs), particularly in ice chemistry, we have also included the chemical network proposed by \citet{belloche2014} and \citet{garrod2017}. We added more than 300 reactions concerning AAN, MCA and MCI into our initial gas-grain chemical network, as listed in Table \ref{tab:gas-phase} and Table \ref{tab:activation-energy}. Table \ref{tab:binding-energy} shows a selection of species' binding energies used in our model. The rate constants of reactions are calculated by the corresponding equations according to \citet{hasegawa1992}, and \citet{ruaud2016} on surface and mantle phases. Since \citet{graedel1982} proposed that the physical assumptions of the process, where diffuse interstellar material collapses into denser clouds, are more suitable for the low metal abundance model, based on a comparison of observations of several molecules and model results, we used the standard oxygen-rich low-metal initial elemental abundances from \citet{graedel1982}, as well as \citet{quan2007}, as listed in Table \ref{tab:initial-abundance}. All abundances are given with respect to the total hydrogen density. The code simulates chemistry in three phases i.e., gas-phase, grain surface, and icy mantle, and also considers various possible exchanges among the different phases via accretion of gas phase species onto grain surfaces, thermal and non-thermal desorption of species from grain surface into the gas phase. The latter includes cosmic-ray desorption, photodesorption via external photons or those via photons produced by cosmic rays \citep{oberg2007}, and reactive desorption using the Rice-Ramsperature-Kessel (RRK) approach \citep{garrod2007}. The surface-mantle and mantle-surface exchange of species is also considered. The diffusion energy to binding energy ratio assumes to be 0.5 for all species on the surface and in ice mantle, respectively. Grains are assumed to be spherical with the radii of 0.1 $\mu$m. The grain density is 3 g cm$^{-3}$ and the gas-to-dust mass ratio is 100. 

\begin{table}
\centering
\caption{Initial elemental abundance compared to total proton density of chemical models.}
\label{tab:initial-abundance}
\begin{tabular}{p{2.5cm}p{2.5cm}}
		\hline
		Species         & Abundance\\
		\hline
		He              & $6.00\times10^{-2}$\\
		N               & $2.14\times10^{-5}$\\
		O               & $1.76\times10^{-4}$\\
		H$_{2}$         & $5.00\times10^{-1}$\\
		C$^{+}$         & $7.30\times10^{-5}$\\
		S$^{+}$         & $8.00\times10^{-8}$\\
		Si$^{+}$        & $8.00\times10^{-9}$\\
		Fe$^{+}$        & $3.00\times10^{-9}$\\
		Na$^{+}$        & $2.00\times10^{-9}$\\
		Mg$^{+}$        & $9.00\times10^{-9}$\\
		P$^{+}$         & $3.00\times10^{-9}$\\
		Cl$^{+}$        & $4.00\times10^{-9}$\\
		F$^{+}$         & $6.69\times10^{-9}$\\
		\hline
\end{tabular} 
\end{table}

AAN has only been detected in the high mass star formation region Sgr B2(N). MCA and MCI have not yet been detected in any sources so far. Therefore, we simulated the physical conditions where AAN was detected, i.e., the hot cores. Correspondingly, we applied hot core models to simulate the formation and destruction processes of AAN, MCA and MCI. Star formation can be divided into two stages based on different periods of evolution. The first stage involves the simulation of protostar growth through the accretion of the outer envelope region, known as the free-fall collapse process or prestellar phase. During this stage, which lasts approximately 1 $\times$ 10$^6$ yr, the temperature remains consistently low, while the density gradually increases over time, resulting in an increase in extinction. The initial gas and grain temperatures are set at 10 K, the initial gas density is 3 $\times$ 10$^{3}$ cm$^{-3}$, and the final collapse density is 1.6 $\times$ 10$^{7}$ cm$^{-3}$. The second stage is the warm-up stage, where the temperature increases according to the formula  $T = T_0 + (T_{max}-T_0)(\Delta t/t_h)^n$ \citep{garrod2006}, with $n$ = 2, t$_h$ = 2 $\times$ 10$^5$ yr, $\Delta t = (t-t_0$), and t$_0$ = 1 $\times$ 10$^5$ yr. After an initial cold phase of 1 $\times$ 10$^5$ yr, the gas and grain temperatures increase from 10 K to the maximum values of T$_{max}$, reaching 150 K and 200 K within a time frame of 2 $\times$ 10$^5$ yr. Once the maximum value is reached, the temperature remains constant. During this stage, it is assumed that the gas and dust temperatures are well coupled. The major physical parameters of hot core models are summarized in Table \ref{tab:physical}. 

\begin{table*}
\centering
\caption{Physical parameters of hot core models}
\label{tab:physical}
\begin{tabular}{p{3cm}p{2.5cm}p{2.5cm}p{2.5cm}p{1.8cm}p{2.5cm}}
\hline
Source (Stage)  &n$_{\rm H}$ (cm$^{-3}$)   &T (K)  &A$_V$ (mag)  &$\zeta$ (s$^{-1}$) &UV factor (Habing)\\
\hline
The freefall collapse$^{a,b}$  &3 $\times10^{3}$ $\rightarrow$ 1.6$\times10^{7}$ &10  &2 $\rightarrow$ 6.109$\times$10$^{2}$ &1.3$\times$10$^{-17}$ &1\\
The warm-up$^{b,c}$  &1.6$\times$10$^{7}$  &10 $\rightarrow$ 150, 200  &6.109$\times$10$^{2}$     &1.3$\times$10$^{-17}$  &1, 10, 100\\
\hline
\multicolumn{6}{l}{Notes. $^a$\citet{garrod2006}, $^b$\citet{bonfand2019}, $^c$ \citet{coutens2018}}
\end{tabular}
\end{table*}

\section{RESULTS}
\label{sect:result}
\subsection{Hot cores}
\begin{figure*}
\centering
\includegraphics[scale=0.6]{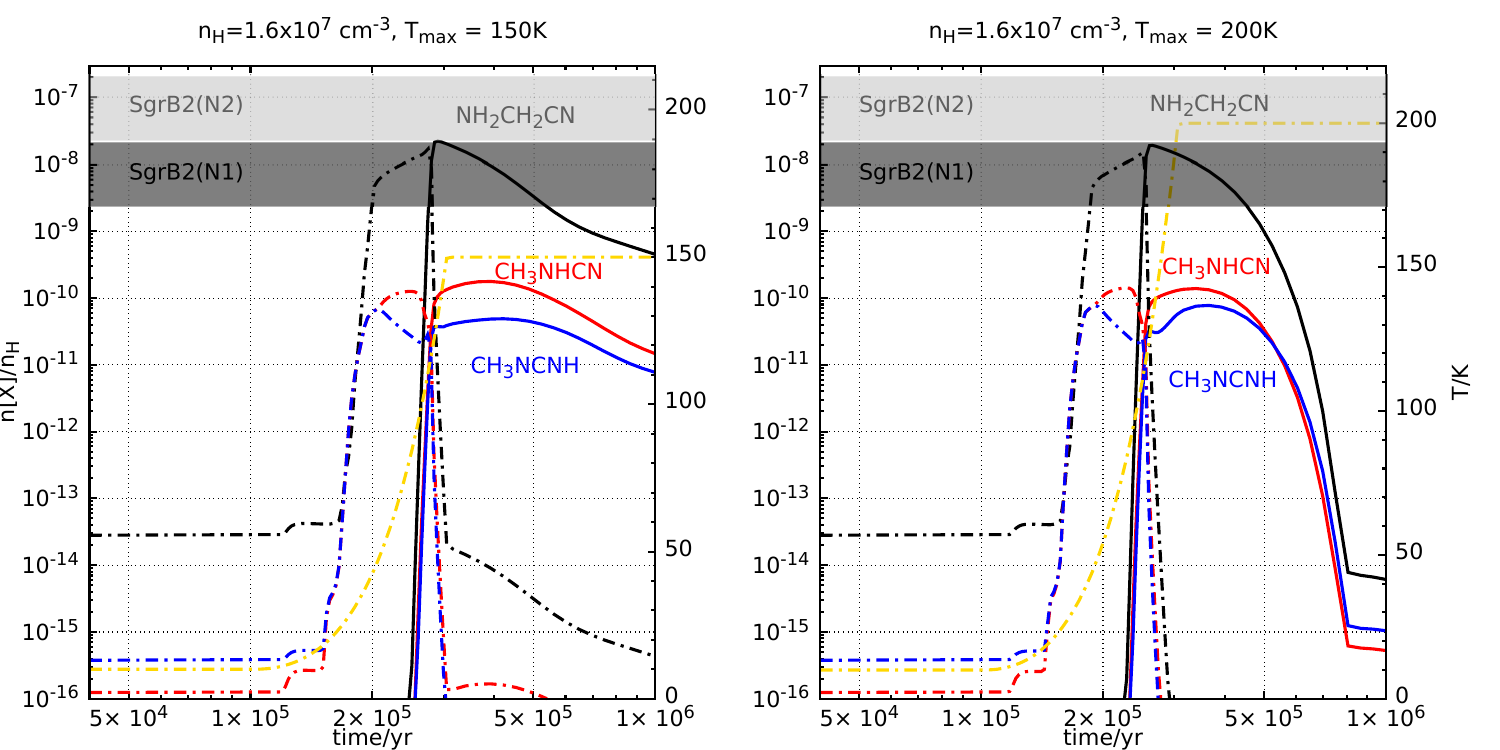}
\caption{The calculated abundances of AAN, MCA and MCI in the gas phase and on grain mantles, including surface and icy mantle, are plotted versus time for warm-up stages in hot core models. The density is 1.6 $\times$ 10$^{7}$ cm$^{-3}$, and the temperatures remain at a constant value of 10 K. During the warm-up stage, T$_{max}$ adopt two values, 150 and 200 K. Solid lines correspond to the gas phase abundances, dotted lines indicate abundances on the grain mantles. The gray and light gray rectangles represent the observed abundance with $\pm$ a factor of 3 uncertainty for AAN in Sgr B2(N1) and Sgr B2(N2), respectively. Golden dotted lines denote temperature profiles.}
\label{fig:prestellar}
\end{figure*}

The results of fractional abundances of AAN, MCA, and MCI molecules with respect to total hydrogen (n$_H$) for two warm-up models are presented in Fig. \ref{fig:prestellar}. These models correspond to different maximum temperatures (T$_{max}$ = 150, 200 K). The physical parameters used in the models are based on the sources of Sgr B2(N) suggested by \citet{bonfand2019}. In Fig. \ref{fig:prestellar}, the gray and light gray rectangles represent the observed abundances of AAN in Sgr B2(N1) and Sgr B2(N2), respectively, with an uncertainty of $\pm$ a factor of 3. The observational results for AAN in Sgr B2(N1) and Sgr B2(N2) were obtained from \citet{melosso2020} and \citet{richard2018}, respectively. For the hot cores Sgr B2(N1) and Sgr B2(N2), we adopted the hydrogen column density of 1.54 $\times$ 10$^{25}$ cm$^{-2}$ and 1.42 $\times$ 10$^{24}$ cm$^{-2}$ respectively, as reported by \citet{bonfand2017, bonfand2019}. The observed abundances of AAN were $\sim$7.14 $\times$ 10$^{-9}$ and $\sim$6.83 $\times$ 10$^{-8}$ in Sgr B2(N1) and Sgr B2(N2), respectively. Here we compare the simulated results with observations toward Sgr B2(N) in Fig. \ref{fig:prestellar}.

The warm-up models produce sufficient abundances of AAN during the warm-up process, reaching maximum temperatures of 150 and 200 K at a hydrogen density of 1.6 $\times$ 10$^{7}$ cm$^{-3}$. At these temperatures, the peak gas phase fractional abundances of AAN can reach 2.23 $\times$ 10$^{-8}$ and 1.93 $\times$ 10$^{-8}$ at the time of 2.89 $\times$ 10$^{5}$ yr and 2.64 $\times$ 10$^{5}$ yr, respectively. Our simulated results are found to be in good agreement with the observed values of Sgr B2(N1) within the time range of (2.74 - 5.63) $\times$ 10$^{5}$ yr and (2.51 - 4.53) $\times$ 10$^{5}$ yr at T$_{max}$ of 150 K and 200 K, respectively. For Sgr B2(N2), the model results are unable to match the observed abundance at these maximum temperatures. In our models, the peak gas phase fractional abundances of MCA can reach 1.77 $\times$ 10$^{-10}$ and 1.39 $\times$ 10$^{-10}$ at the time of 3.92 $\times$ 10$^{5}$ yr and 3.40 $\times$ 10$^{5}$ yr for T$_{max}$ = 150 K and 200 K, respectively. For MCI, the peak values are 4.88 $\times$ 10$^{-11}$ and 7.74$\times$ 10$^{-11}$ at the time of 4.22 $\times$ 10$^{5}$ yr and 3.65 $\times$ 10$^{5}$ yr for T$_{max}$ = 150 K and 200 K, respectively.

Gas phase AAN mainly originate from the surface through thermal desorption processes. A significant amount of surface AAN desorb into the gas phase at temperature up to approximately 100 K, which is lower than the experimental results $\geq$ 190 K from \citet{borget2012}. We utilized a binding energy of 5480 K for all three isomers \citep{garrod2013}, as presented in Table \ref{tab:binding-energy}. As illustrated in Fig. \ref{fig:prestellar}, it can be seen that the abundances of the three isomers increase significantly as the temperature rises. This can be attributed to the availability of sufficient thermal energy to enable reactants to overcome the diffusion barriers and react with each other on the grain surface, leading to the formation of products that evaporated into gas phase at a temperature of approximately 100 K. 

During the warm-up stage, there are several major formation reactions involved in the production of AAN on the grain surface, including NH$_2$ + H$_2$CCN, HNC + CH$_2$NH and NHCH$_2$CN + H. NHCH$_2$CN is formed through the reactions of NH + H$_2$CCN and CN + CH$_2$NH on the grain surface. The first formation route agrees with the modeling result of \citet{garrod2013}, while the validity of the third route has been supported by \citet{belloche2009}. Although, the gaseous AAN primarily originates from the thermal desorption process of grain surface, the peak abundance of AAN in the gas phase exceeds its peak abundances on grain surface in our models. This can be explained as follows: When AAN reaches its peak abundance in the gas phase, the reaction of CH$_2$NH + HNC on grain surface is the major formation route for AAN. Simultaneously, CH$_2$NH is formed through two pathways: the thermal desorption process and the reaction of NH + CH$_3$ in the gas phase. This results in a significant increase in the abundance of CH$_2$NH in the gas phase. Additionally, more CH$_2$NH accretes on the grain surface according to the equation for the rate of accretion of a gas-phase species \citep{semenov2010}. Subsequently, more gaseous AAN is produced as CH$_2$NH reacts with HNC to form AAN on the grain surface, followed by the thermal desorption process. As a result, the gas phase abundances of AAN exceed its grain surface phase abundances. On the other hand, MCA and MCI are primarily formed through the reactions of CH$_3$NCN + H on the grain surface and the electronic recombination of ion C$_2$H$_5$N$_2^+$ in the gas phase. CH$_3$NCN is produced through the hydrogenation reaction of CH$_2$NCN on the grain surface. In our previous modeling work, we found that CH$_2$NCN is present in relatively low abundance \citep{zhang2020}. Recently, it has been detected in the Galactic Center G+0.693-0.027 molecular cloud \citep{san2024}. The C$_2$H$_5$N$_2^+$ ion is mainly formed via the ion-neutral reaction of AAN. Eventually, as the temperature reaches 100 K, the isomers efficiently desorb from the grain surface into the gas phase. This desorption behavior during the warm-up stage aligns with the mechanism proposed by \citet{garrod2006} for studying complex organic molecules in ISM. However, after reaching their peak abundances, the three isomers abundances subsequently decline due to active destruction reactions in the gas phase, including reactions with positive ions such as H$^+$, H$_3^+$, C$^+$, He$^+$ and H$_3$O$^+$, as well as the free radical OH. As illustrated in Fig. \ref{fig:prestellar}, there are rapidly decreasing abundance for these isomers at T$_{max}$ = 200 K due to the higher temperature results in a higher destruction rate comparing to T$_{max}$ = 150 K.

In our models, the difference in abundances between AAN and MCA, as well as between MCA and MCI, is around 1 to 2 orders of magnitude at T$_{max}$ = 150 K. However, at T$_{max}$ = 200 K, the difference between MCA and MCI is less than one order of magnitude. Thus, the discrepancy in the abundances of three isomers is significant at different maximum temperatures. We speculated that MCA and MCI could be potential interstellar molecules and may be detectable in the sources where AAN is abundant, using available spectral data. Since a large part of MCA and MCI are derived from the ion-neutral reactions of AAN and subsequently by the electronic recombination of ion C$_2$H$_5$N$_2^+$ in the gas phase. However, due to its higher structural instability and rapid chemical destructive rate, MCI is more challenging to detect in the ISM compared to MCA. 

\begin{figure*}
\centering
\includegraphics[scale=0.6]{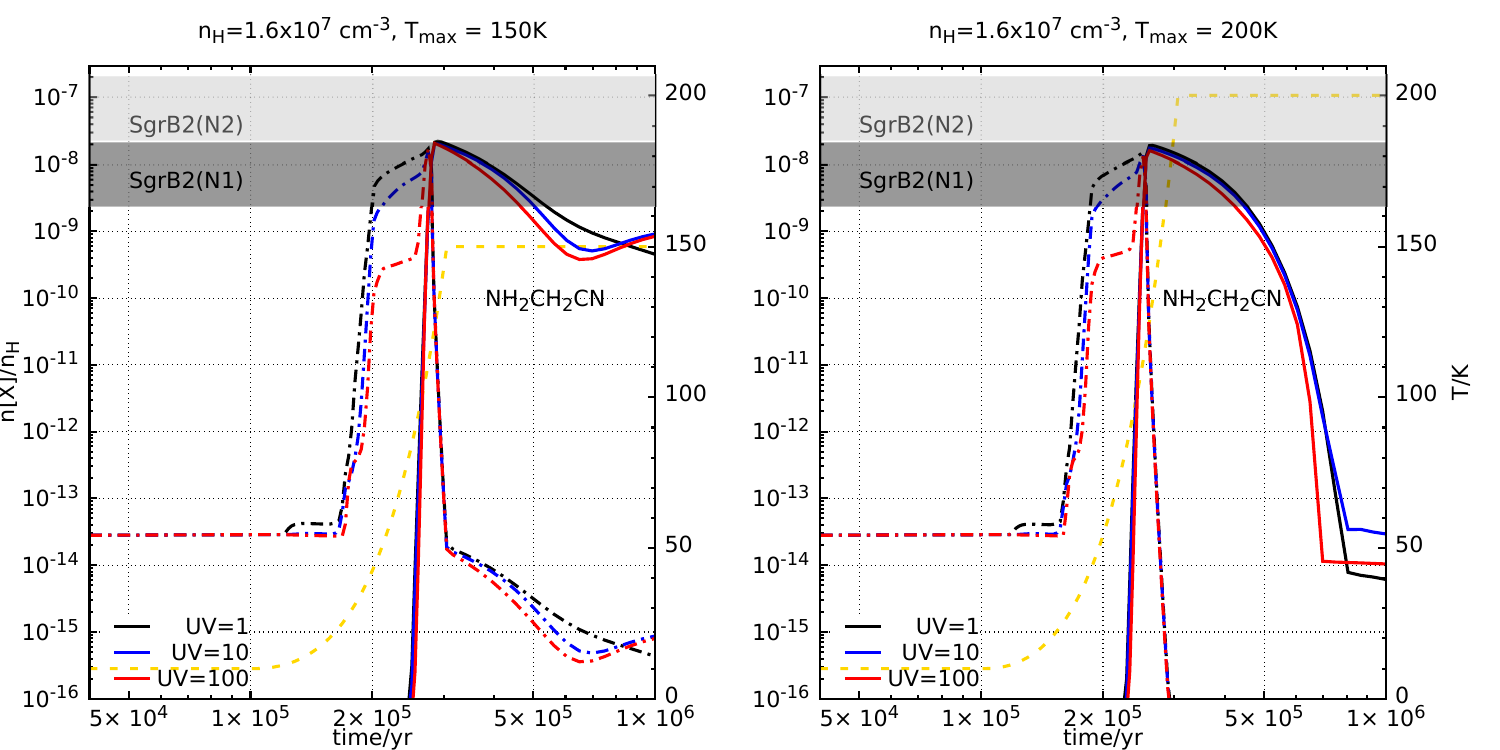}
\caption{The calculated abundances of AAN in the gas phase and on the grain mantle, are plotted versus time for warm-up models with the gas density of 1.6 $\times$ 10$^{7}$ cm$^{-3}$. The left panel denotes T$_{max}$ of 150 K with different UV factors, and the right panel responds T$_{max}$ of 200 K. Solid lines correspond to the gas phase abundances, dotted lines indicate abundances on the grain mantles. The gray and light gray rectangles represent the observed abundance with $\pm$ a factor of 3 uncertainty for AAN in Sgr B2(N1) and Sgr B2(N2), respectively. Golden dotted lines denote temperature profiles.}
\label{fig:prestellar_uv}
\end{figure*}

\begin{figure*}
\centering
\includegraphics[scale=0.6]{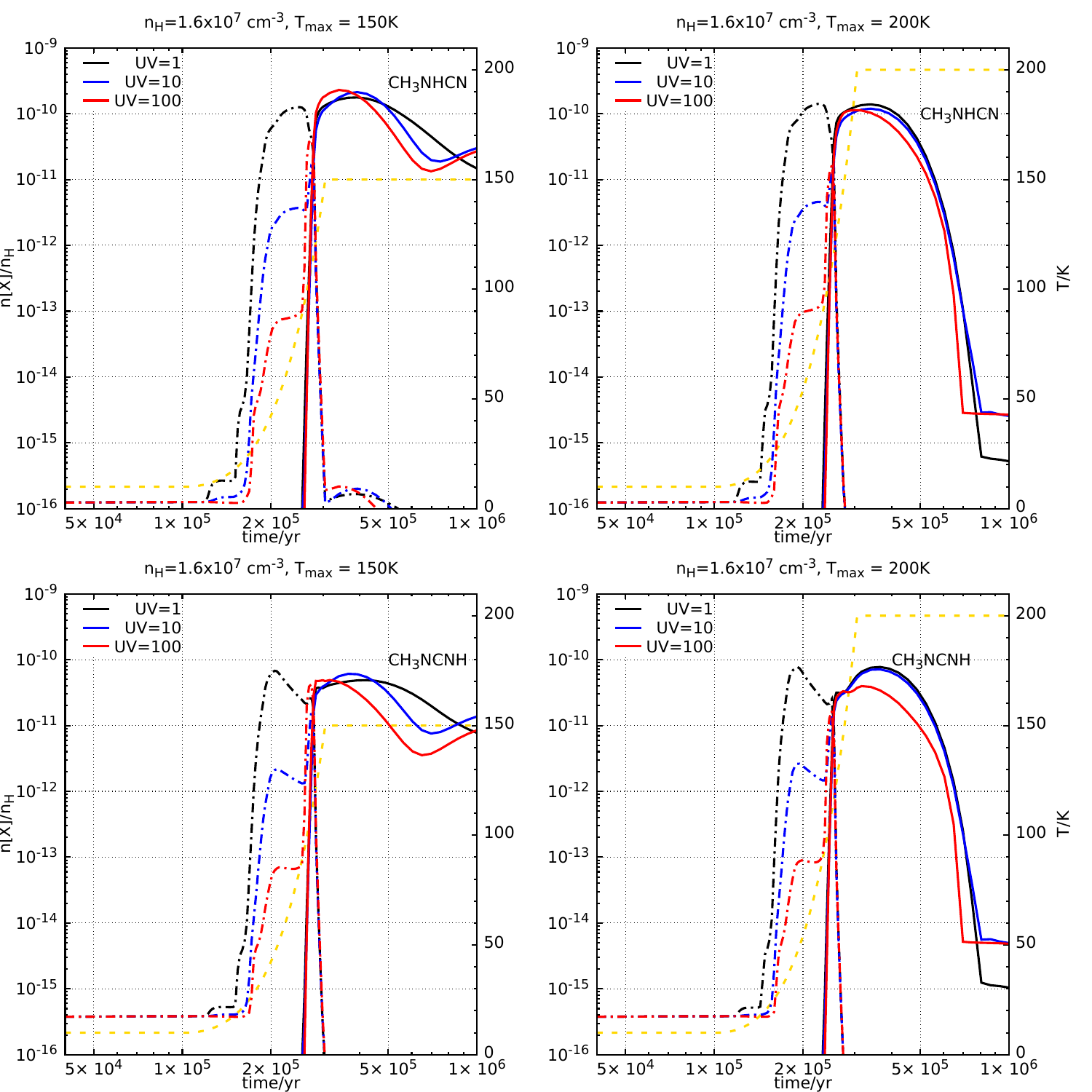}
\caption{The calculated abundances of MCA and MCI in the gas phase and on the grain mantle, are plotted versus time for warm-up models with the gas density of 1.6 $\times$ 10$^{7}$ cm$^{-3}$. The left panel denotes T$_{max}$ = 150 K with different UV factors, and the right panel responds T$_{max}$ = 200 K. Solid lines correspond to the gas phase abundances, dotted lines indicate abundances on the grain mantles. Golden dotted lines denote temperature profiles.}
\label{fig:prestellar_uv2}
\end{figure*}

In high mass star formation regions, there is a significant presence of UV radiation. Therefore, in order to further simulate the physical conditions in Sgr B2(N), we utilized two higher values of UV factor, 10 and 100 times larger than the normal value for models with a hydrogen density of 1.6 $\times$ 10$^{7}$ cm$^{-3}$. It should be noted that only the warm-up stage in the models adopted the high UV factors. The results of these simulations are shown in Fig. \ref{fig:prestellar_uv}. It can be seen that fractional abundances of AAN are lower comparing to the results obtained from the standard UV factor computations, for both T$_{max}$ of 150 and 200 K. The stronger UV radiations lead to high photodissociation rate for the three isomers. The peak abundance of AAN drops from 2.23 $\times$ 10$^{-8}$ to 2.03 $\times$ 10$^{-8}$ at a time of 2.89 $\times$ 10$^{5}$ yr as the UV factor increases from 1 to 10 for T$_{max}$ of 150 K. For the case of UV factor of 100, it has the same trend as the UV factor of 10, and the peak abundance of AAN deduced to 1.98 $\times$ 10$^{-8}$ at the same time. For T$_{max}$ = 200 K, the abundance of AAN decreases from 1.93 $\times$ 10$^{-8}$ to 1.73 $\times$ 10$^{-8}$ at a time of 2.64 $\times$ 10$^{5}$ yr as the UV factor increases from 1 to 10. When a UV factor of 100 is applied, the abundance of AAN further reduced to 1.55 $\times$ 10$^{-8}$. A higher temperature leads to a lower abundance of AAN compared to T$_{max}$ = 150 K, as the higher temperature results in a higher destruction rate. However, the high UV factor only slightly reduces the abundance of AAN compared to the simulated result using the standard UV factor. There are also experimental findings that suggest species containing nitriles survive longer than the corresponding amino acids glycine when exposed to UV photolysis in the solid state at 15 K \citep{bernstein2004}. The simulated abundances of AAN also agree well with the observations within a certain time range, but the time range is shorter than using the standard UV factor, as shown in Fig. \ref{fig:prestellar_uv}. Similarly, MCA exhibits a similar trend at high UV factors to AAN, because it also contain nitrile group. Therefore, higher UV factors have minimal effects on the differences in their abundance magnitudes. In contrast, MCI shows a different behavior from AAN and MCA. Its abundance is greatly influenced by UV radiation due to the absence of nitrile group, making it susceptible to UV photodissociation, as shown in Fig. \ref{fig:prestellar_uv2}.

\subsection{The chemical relations among AAN isomers and other species}
\begin{figure*}
\centering
\includegraphics[scale=0.6]{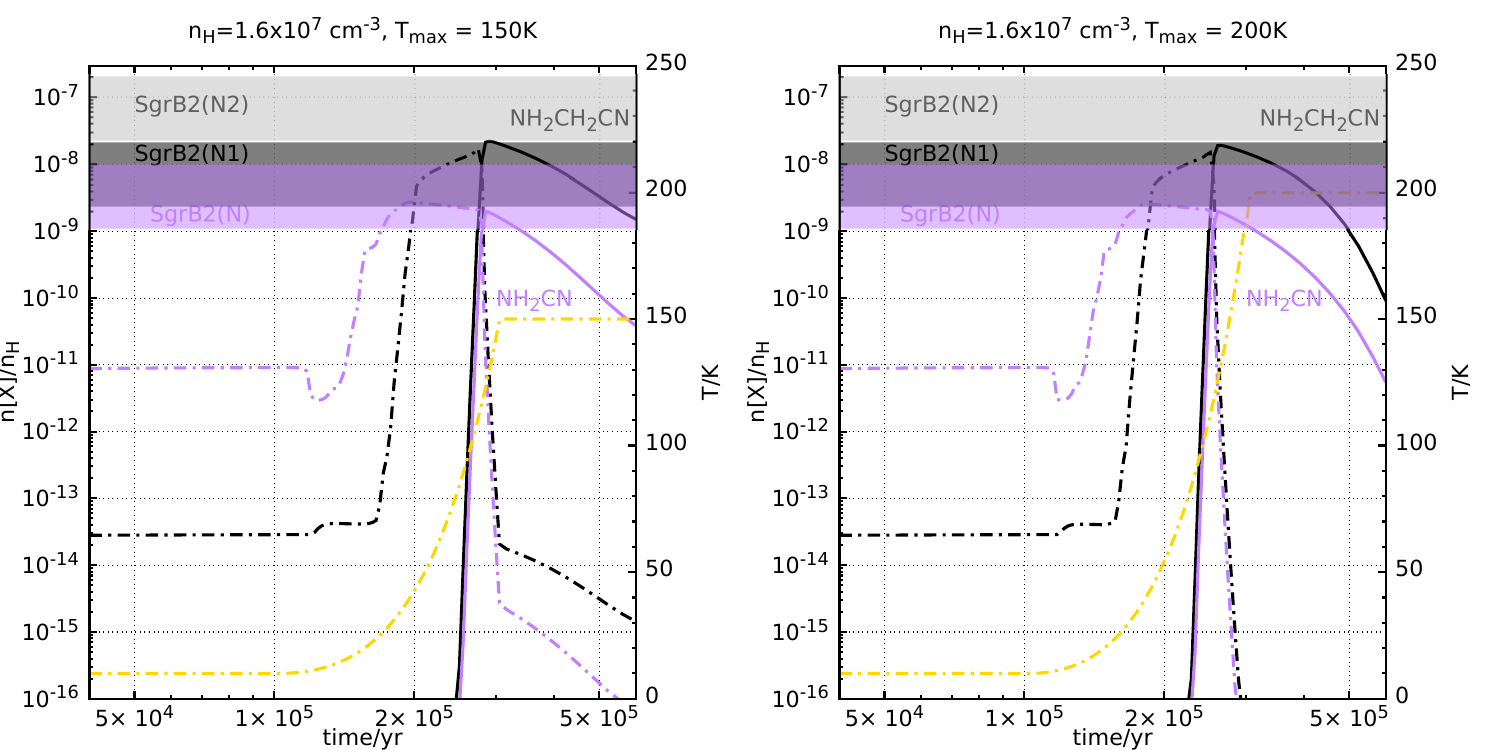}
\caption{The calculated abundances of AAN and NH$_2$CN in the gas phase and on grain mantles, including surface and icy mantle, are plotted versus time for warm-up stages in hot core models. The density is 1.6 $\times$ 10$^{7}$ cm$^{-3}$, and the temperatures remain at a constant value of 10 K. During the warm-up stage, T$_{max}$ = 150 K, 200 K. Solid lines correspond to the gas phase abundances, dotted lines indicate abundances on the grain mantles. The gray, light gray and purple rectangles represent the observed abundance with $\pm$ a factor of 3 uncertainty for AAN in Sgr B2(N1) and Sgr B2(N2) and for NH$_2$CN in Sgr B2(N), respectively. Golden dotted lines denote temperature profiles.}
\label{fig:ratio_nh2cn}
\end{figure*}

AAN is a compound that contains both an amino and a nitrile functional group, making it an amino nitrile. In astrobiology, many bifunctional building blocks can undergo cyclization reactions, allowing rapid advance toward prospective heterocyclic units. AAN can synthesize adenine through two-step chemical reactions \citep{vasconcelos2020}. The simplest amino nitrile is NH$_2$CN \citep{cleaves2011}, which can also form adenine via a series of chemical reactions \citep{merz2014}. NH$_2$CN has been detected in many different types of star formation regions, as listed in \citet{zhang2023}. Here we calculated the ratio of NH$_2$CN to AAN and found the same value of 0.1 when their abundances both reach their peaks at T$_{max}$ = 150 K and 200 K corresponding to the time of  2.89 $\times$ 10$^{5}$ yr and 2.64 $\times$ 10$^{5}$ yr. as shown in Fig. \ref{fig:ratio_nh2cn}. The column density of NH$_2$CN was 5.13 $\times$ 10$^{16}$ cm$^{-2}$ with the rotational temperature of 150 K towards the Sgr B2(N) \citep{belloche2013}. The column density of AAN is 9.7 $\times$ 10$^{16}$ cm$^{-2}$ in the same region with the same rotational temperature \citep{richard2018}. The observational ratio is 0.53, which is very close to our simulated ratio.

During warm-up stage, both AAN and NH$_2$CN undergo similar chemical formation processes. Initially, they are formed via free radicals reactions on the grain surface at a relative low temperature. Subsequently, they desorb from the grain surface and enter the gas phase through thermal desorption when the temperature reaches a high value. Therefore, their were mainly detected in star-formation regions. However, NH$_2$CN is observed in a wider variety of sources due to its smaller size relative to AAN and its more facile formation in ISM. Generally, the smaller molecules of the same family are more abundant than the larger ones. To the contrary, the abundances of AAN, despite its higher molecular mass, are higher than those of NH$_2$CN. In our models, we found that NH$_2$CN has higher desorption energy than that of AAN, which may be a contributing factor to this unexpected outcome. However, this does not alter the results when we exchange the desorption energies of NH$_2$CN and AAN. Therefore, we speculated that the reactants of AAN are easy to form and have high abundances comparing to those of NH$_2$CN. APN is another member of the AAN family, which is the next stage in complexity with one additional CH$_2$ group. \citet{richard2018} conducted a search for APN towards the Sgr B2(N), but their attempts were unsuccessful. Based on our present work, we believed that the formation of APN in the ISM is more challenging compared to AAN.

\begin{figure*}
\centering
\includegraphics[scale=0.6]{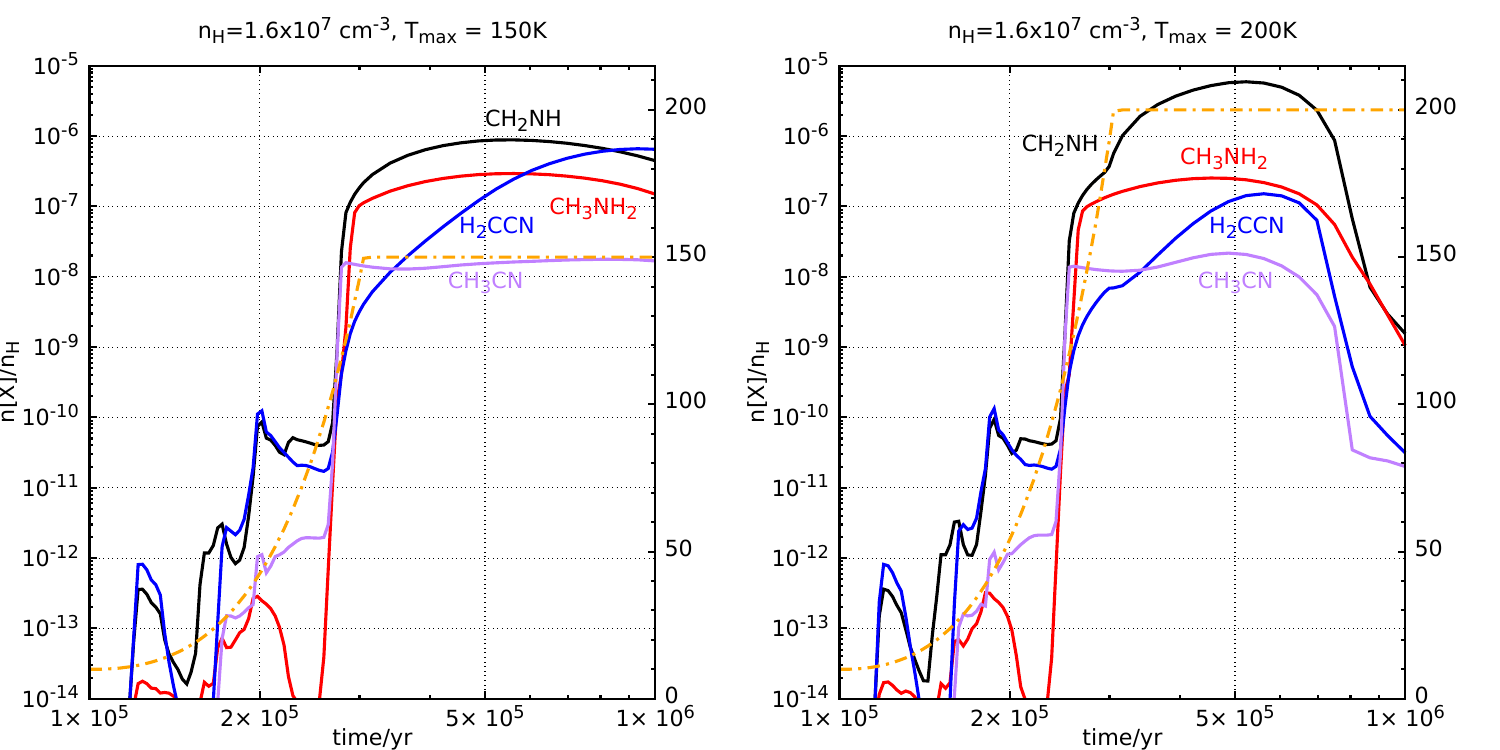}
\caption{The calculated abundances of CH$_2$NH, CH$_3$NH$_2$, H$_2$CCN and CH$_3$CN in the gas phase are plotted versus time for warm-up stages in hot core models. The density is 1.6 $\times$ 10$^{7}$ cm$^{-3}$, and the temperatures remain at a constant value of 10 K. During the warm-up stage, T$_{max}$ = 150 and 200 K. Solid lines correspond to the gas phase abundances of CH$_2$NH, CH$_3$NH$_2$, H$_2$CCN and CH$_3$CN, Dotted lines denote temperature profiles.}
\label{fig:prestellar_react}
\end{figure*}

According to \citet{garrod2013}, AAN is primarily formed through the reaction of NH$_2$ + H$_2$CCN on grain surface. However, in our model, this reaction is not the sole major formation pathway. Additionally, their proposal of H$_2$CCN originating from CH$_3$CN differs from our results. In our study, H$_2$CCN is mainly produced via the reaction of CN + CH$_3$ (Reaction (6)), as described by \citet{loison2014}. It could also be formed through the electronic recombination of ion CH$_3$CNH$^+$ in the gas phase (Reaction (7)). The branching ratio of the channel leading to H$_2$CCN is approximately 24 percent \citep{loison2014}. Furthermore, CH$_3$CNH$^+$ is primarily generated through the reaction of HCN + CH$_3^+$ in the gas phase rather than the protonated CH$_3$CN. On the other hand, CH$_3$CN is mainly produced through the electronic recombination of ion CH$_3$CNH$^+$, which branching ratio of this channel giving CH$_3$CN is approximately 39 percent (Reaction (8))\citep{loison2014}. On grain surface, it is mainly formed by the hydrogenation of H$_2$CCN. Both H$_2$CCN and CH$_3$CN are destroyed by ion-neutral reactions involving H$_3^+$ in gas phase, as well as by the hydrogenation reaction on grain surface. Therefore, we concluded that AAN and CH$_3$CN weakly compete for the utilization of H$_2$CCN as a precursor on grain surfaces. As shown in Fig. \ref{fig:prestellar_react}, the peak abundance of CH$_3$CN are 1.77 $\times$ 10$^{-8}$ at T$_{max}$ = 150 K, which are one order of magnitude higher than the observed towards SgrB2(N) \citep{belloche2013}. The peak abundance of H$_2$CCN are 6.6 $\times$ 10$^{-7}$ at T$_{max}$ = 150 K. However, it has not been detected towards the SgrB2(N) so far. Here we only discussed the results for a temperature of T$_{max}$ = 150 K, as T$_{max}$ = 200 K shows a similar trend but with a higher destruction rate. 
\protect\\

\noindent CN + CH$_3$ $\rightarrow$ H$_2$CCN + H, \hfill(6)  \protect\\

\noindent CH$_3$CNH$^+$ + e$^-$ $\rightarrow$ H$_2$CCN + H + H, \hfill(7)  \protect\\

\noindent CH$_3$CNH$^+$ + e$^-$ $\rightarrow$ CH$_3$CN + H. \hfill(8) 
\protect\\

CH$_2$NH and CH$_3$NH play a significant role in the formation of AAN and MCA, making them crucial components for the detection of AAN and MCA in the ISM. Additionally, CH$_2$NH is an important prebiotic molecule that has been suggested as a possible precursor for glycine and its nitrile AAN \citep{hoyle1976, dickerson1978} and has been detected in several Galactic objects \citep{godfrey1973, dickens1997, qin2010, halfen2013, suzuki2016}. However, CH$_3$NH has not been detected in any sources thus far. In our models, CH$_2$NH is formed through both the neutral–neutral reaction of NH + CH$_3$ (Reaction (9)) and the electronic recombination of CH$_3$NH$_3^+$ ion in gas phase (Reaction (10)), as well as the hydrogenation reaction of H + H$_2$CN on the grain surface. It is destroyed through ion-neutral reactions with H$_3^+$ in gas phase and through the hydrogenation reaction to form CH$_3$NH and CH$_2$NH$_2$ on the grain surface, with activation energies of 2134 K and 3170 K, respectively. This hydrogenation reaction is the major pathway for the formation of CH$_3$NH. CH$_3$NH can be formed in gas phase through the reaction of CN + CH$_3$NH$_2$, albeit as a minor contribution. CH$_3$NH is ultimately transformed into CH$_3$NH$_2$ on the grain surface, which is also believed to be a potential interstellar precursor to the amino acid glycine, e.g., \citet{bossa2009, lee2009}. \citet{halfen2013} suggested that CH$_2$NH and CH$_3$NH$_2$ are formed via different chemical processes in Sgr B2 based on the rotational temperatures and distributions observed in Sgr B2. They also proposed that CH$_2$NH exists in both a colder, foreground gas and a warmer, background cloud, while CH$_3$NH$_2$ is present in much warmer gas compared to CH$_2$NH. This is because CH$_3$NH$_2$ is primarily formed through the reaction of CH$_3$NH + H (major pathway) and CH$_2$NH$_2$ + H (minor pathway) on the grain surface, and when the temperature rises, it evaporates through thermal processes. On the other hand, CH$_2$NH is formed both in the gas phase through Reaction (9) and to a lesser extent on grain surface through the reaction of H + H$_2$CN. Therefore, it can exist in regions that are not as warm \citep{halfen2013}. Fig. \ref{fig:prestellar_react} shows that the peak abundance of CH$_2$NH is 8.83 $\times$ 10$^{-7}$ at T$_{max}$ = 150 K, which is one order of magnitude higher than the observation in SgrB2(N) \citep{belloche2013, suzuki2016}, similar to the results of \citet{suzuki2016}. The peak abundance of CH$_3$NH$_2$ is 2.93 $\times$ 10$^{-7}$ at T$_{max}$ = 150 K, which is overproduced by one order of magnitude compared to the observation from \citet{belloche2013}. For T$_{max}$ = 200 K, there is a higher production of CH$_2$NH molecules compared to T$_{max}$ = 150 K, but for CH$_3$NH$_2$, the trend is similar. Additionally, both exhibit a higher molecular destruction rate. \protect\\ 

\noindent NH + CH$_3$ $\rightarrow$ CH$_2$NH + H, \hfill(9)  \protect\\

\noindent CH$_3$NH$_3^+$ + e$^-$ $\rightarrow$ CH$_2$NH + H$_2$ + H. \hfill(10)  \protect\\

\begin{figure*}
\centering
\includegraphics[scale=0.6]{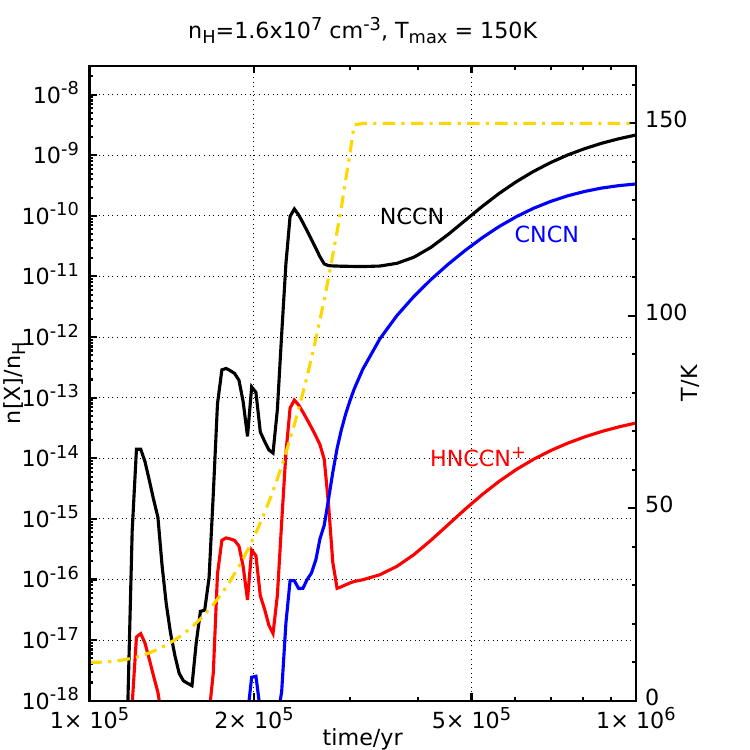}
\caption{The calculated abundances of NCCN, CNCN and NCCNH$^+$ in the gas phase are plotted versus time for warm-up stages in hot core models. The density is 1.6 $\times$ 10$^{7}$ cm$^{-3}$, and the temperatures remain at a constant value of 10 K. During the warm-up stage, T$_{max}$ = 150 K. Solid lines correspond to the gas phase abundances of NCCN, CNCN and NCCNH$^+$, Dotted lines denote temperature profiles.}
\label{fig:nccn}
\end{figure*}

Cyanogen (NCCN) is another important molecules that may possible react with AAN to form C$_4$H$_4$N$_4$, then reacts with HCN to produce adenine, which is one of purine of DNA and RNA nucleobases \citep{vasconcelos2020}. Additionally, NCCN is the simplest member of the series of dicyanopolyyne,(N$\equiv$C–(C$\equiv$C)$_n$–C$\equiv$N), which are believed to be potentially abundant interstellar molecules \citep{kolos2000,petrie2003}. However, NCCN is non-polar molecule and cannot be directly observed. Its presence in interstellar space was indirectly detected for the first time through the detection of the protonated form (NCCNH$^+$) in the dense clouds L483 and TMC-1 \citep{Agundez2015}. That was further confirmed by the detection of the metastable and polar isomer, isocyanogen (CNCN) \citep{Agundez2018}. The fractional abundances of  NCCNH$^+$ and CNCN in L483 and TMC-1 are fairly low, ranging from (1 - 10) $\times$ 10$^{-12}$ and (5 - 9) $\times$ 10$^{-11}$, respectively \citep{Agundez2015,Agundez2018}. \citet{Agundez2018} suggested that NCCN could be fairly abundant, with an abundance in dark clouds ranging from 10$^{-9}$ – 10$^{-7}$. In our hot core models, the peak abundances of NCCN, CNCN and NCCNH$^+$  can reach 4.38 $\times$ 10$^{-9}$, 3.50 $\times$ 10$^{-10}$ and 1.96 $\times$ 10$^{-13}$, respectively, as shown in Fig. \ref{fig:nccn}. The simulated ratio of NCCNH$^+$ to NCCN is 9.25 $\times$ 10$^{-5}$, with NCCNH$^+$ reaching its peak abundance at the time of 2.05 $\times$ 10$^{6}$ yr. These results are consistent with the chemical models predicted by \citet{Agundez2015}. The major formation pathways of NCCN involve the reactions of CN + HNC, N + C$_3$N in the gas phase. In the future, NCCN could potentially be indirectly observed by detecting the NCCNH$^+$ ion, or inferred by detecting its higher-energy polar isomer, CNCN, in hot cores. Based on our model results, it is possible that NCCN and AAN react to form C$_4$H$_4$N$_4$, as they have relatively high abundances in hot core models. However, the reaction need to be verified through theoretical calculations or laboratory experiments. Here we only show the results for a temperature of T$_{max}$ = 150 K, as T$_{max}$ = 200 K shows a similar trend. 

\section{CONCLUSIONS}
\label{sect:conclu}
In this work, we conducted chemical simulations to calculate the evolution of NH$_2$CH$_2$CN and its isomers in the ISM. The main results can be summarized as follows: 

1. With the updated gas-grain chemical network of the related species along with AAN, our models predicted a peak gas phase abundance of AAN on the order of 10$^{-8}$, which is consistent with observations towards Sgr B2(N).

2. Our modeling results suggested the formation processes of AAN and its isomers. We found that AAN is formed on grains through the reaction of NH$_2$ + H$_2$CCN, HNC + CH$_2$NH and NHCH$_2$CN + H at early evolutionary stages. Subsequently, it thermally evaporates into the gas phase when temperature reaches 100 K and destroyed by positive ions (e.g., H$^+$, H$_3^+$) and free radical OH in the gas phase. Its isomer MCA and MCI are formed though the hydrogenation reaction of CH$_3$NCN on the grain surfaces and the electron recombination reactions of the ion C$_2$H$_5$N$_2^+$ in the gas phase.

3. Strong UV radiation had minimal effect on the abundances of AAN and MCA, resulting in changes of no more than one order of magnitude. This finding confirms that species containing nitriles survive longer than the corresponding acids when exposed to UV photolysis in the solid state.

4. AAN and NH$_2$CN are members of the amino nitriles family,  as they contain both an amino and a nitrile functional group. They have similar formation processes, occurring through formation on grain surfaces and subsequent thermal evaporation into the gas phase. The ratio of NH$_2$CN to AAN is 0.1, which is close to the observed value of 0.53.  

5. NCCN, the simplest member of the dicyanopolyyne series, can be formed through reactions of CN + HNC, N + C$_3$N in the gas phase. Our models show that its abundance can reach the order of 10$^{-9}$. Therefore, it is possible that NCCN and AAN react with each other to eventually form adenine in hot cores. This process should be investigated through theoretical calculations or laboratory experiments to understand the validity of grain surface formation of adenine in future work.

\section*{Acknowledgements}

This work was supported by the National Natural Science Foundation of China under grant 12203091, 12373026, 12173075, the Natural Science Foundation of Xinjiang Uygur Autonomous Region (2022D01A156), the "Tianchi Doctoral Program 2021", the National Key R\&D Program of China (No.2022YFA1603103), Youth Innovation Promotion Association CAS.

\section*{DATA AVAILABILITY}
The datasets generated during this study are available in the article and from the corresponding author upon request.



\bibliographystyle{mnras}
\bibliography{bibtex} 








\bsp	
\label{lastpage}
\end{document}